\documentclass[useAMS,usenatbib]{mn2e}

\usepackage{graphicx}
\usepackage{dcolumn}
\usepackage{bm}
\usepackage{amsmath}
\newcounter{ichi}
\newcounter{ni}
\newcounter{san}
\newcounter{yon}
\def\gtrsim{\mathrel{\hbox{\rlap{\hbox{\lower4pt\hbox{$\sim$}}}\hbox{$>$}}}}
\def\lesssim{\mathrel{\hbox{\rlap{\hbox{\lower4pt\hbox{$\sim$}}}\hbox{$<$}}}}

\title[Hadronic emission from interaction-powered supernovae]
{Probing cosmic-ray ion acceleration with radio-submm and gamma-ray emission from interaction-powered supernovae}
\author[K. Murase et al.]
{Kohta Murase$^{1}$, Todd A. Thompson$^{2,3}$, and Eran O. Ofek$^{4}$
\\
$^{1}$Hubble Fellow --- Institute for Advanced Study, Princeton, New Jersey 08540, USA\\
$^{2}$Center for Cosmology and AstroParticle Physics, The Ohio State University, Columbus, Ohio 43210, USA\\
$^{3}$Department of Astronomy, The Ohio State University, Columbus, Ohio 43210, USA\\
$^{4}$Benoziyo Center for Astrophysics, Weizmann Institute of Science, Rehovot 76100, Israel\\
}
\begin{document}

\date{\today}
\pagerange{\pageref{firstpage}--\pageref{lastpage}} \pubyear{2009}
\maketitle
\label{firstpage}

\begin{abstract}
The optical and near-IR emission from some classes of supernovae (SNe), including Type IIn and possibly some super-luminous SNe, is likely powered by a collision between the SN ejecta and dense circumstellar material (CSM).  We argue that for a range of CSM masses and their radii, a collisionless shock can form, allowing for efficient cosmic-ray (CR) acceleration.  We show that $pp$ collisions between these newly accelerated CRs and the CSM leads to not only gamma rays but also secondary electrons and positrons that radiate synchrotron photons in the high-frequency radio bands.  Our estimates imply that various facilities including the Jansky Very Large Array  (VLA) and the Atacama Large Millimeter/submillimeter Array (ALMA) may observe such SNe at Gpc distances by followup observations in months-to-years, although the detectability strongly depends on the CSM density as well as observed frequency.  Detecting this signal would give us a unique probe of CR acceleration at early times, and even non-detections can put interesting limits on the possibility of CR ion acceleration.  Following our previous work, we also show that GeV gamma rays can escape from the system without severe attenuation, encouraging point-source and stacking analyses with {\it Fermi}.  We provide recipes for diagnosing interaction-powered SN scenario with multi-messenger (neutrino and gamma-ray) observations.  
\end{abstract}

\begin{keywords}
non-thermal---supernovae
\end{keywords}

\section{Introduction}
Blind surveys for optical transients have revealed a class of super-luminous supernovae (SL) SNe that may in some cases be powered by a collision between the SN ejecta and a massive shell or wind of circumstellar material (CSM)~\citep[e.g.,][]{fa73,ofe+07,sm07,qui+11}.  Examples include SN 2003ma~\cite{res+11}, 2006gy~\citep[e.g.,][]{ofe+07,smi+07,smi+10}, and 2008es~\cite{mil+09}, among others.  As a consequence of the collision with the CSM, a significant fraction of the kinetic energy is converted into radiation via shock dissipation, which is responsible for the observed emission (see Figure 1).  

The rate of SLSNe with absolute magnitude $M<-21$  is order of $\sim{10}~{\rm Gpc}^{-3}~{\rm yr}^{-1}$, $\sim0.01$\% of the normal core-collapse SN rate~\cite{gal12}, but some normal-luminosity SNe such as SN 2005ip~\cite{smi+09}, 2006jc~\cite{imm+08,smi+08}, 2008iy~\cite{mil+10} and PTF 09uj~\cite{ofe+10}, which may also be powered by ejecta-CSM interactions, are more common~\cite{qui+13}.  Finally, recent observations of SN 2009ip and 2010mc suggest that the CSM eruption is timed to occur months-to-years before the core collapse~\citep{mau+13,pas+13,pri+13,ofe+13,mar+13,ofe+13c}.

Interaction-powered SNe may be efficient cosmic-ray (CR) accelerators, where one can expect that the diffusive shock acceleration mechanism operates at the forward and reverse shocks by a collision between the SN ejecta and CSM.  For a range of CSM parameters (mass and shock dissipation radius), the shock is radiation-mediated --- the Thomson optical depth is larger than $c/V_s$~\cite{wea76,kat+10}, where $V_s$ is the shock velocity --- and efficient CR acceleration is not expected because the CR collisionless mean free path is much shorter than the deceleration length.  However, as the shock propagates in the CSM, photons can stream out ahead of the shock, and photon energy can no longer support the shock (i.e., shock breakout).  After the breakout, for wind-like CSM profiles, the shock will become collisonless and CR acceleration can be efficient~\cite{mur+11,kat+11,kas+13}.  Recently, Murase et al. (2011) considered a collision between the SN ejecta with a CSM shell and found that CR protons may be accelerated, and furthermore that the protons may experience strong pionic losses via inelastic $pp$ collisions, producing gamma-rays and neutrinos.  Thus, interaction-powered SNe may be interesting CR accelerators and high-energy/multi-messenger emitters.  In this work, we continue our study of the possibility of the non-thermal emission from the shock interaction of a SN embedded in a dense CSM.  In particular, we focus on the secondary electrons and positrons expected from the same $pp$ collisions that give rise to neutrinos and gamma rays.  Importantly, we show that these secondaries can emit detectable synchrotron radiation at high-frequency radio wavelengths including mm/submm and FIR bands.

In Section 2, we review the shock physics and the potential for CR acceleration in interaction-powered SNe, providing a much more detailed discussion than Murase et al. (2011).  Section 3 gives a brief discussion of the high-energy emission expected, and recipes that connect the observed optical emission to the non-thermal signatures are provided in Appendix A.  In Section 4 we discuss high-frequency radio diagnostics.  For a range of CSM parameters, we show that secondary leptons from $pp$ interactions should radiate synchrotron at $\sim3-3000~{\rm GHz}$, and with fluxes of $\sim0.01-0.1$~mJy at distances of hundreds of Mpc.  In Section 5, we summarize our results.  

Throughout this work, we use the notation $Q=Q_x{10}^{x}$ in CGS unit unless we give notice. 

\begin{figure}
\includegraphics[width=0.75\linewidth,angle=270]{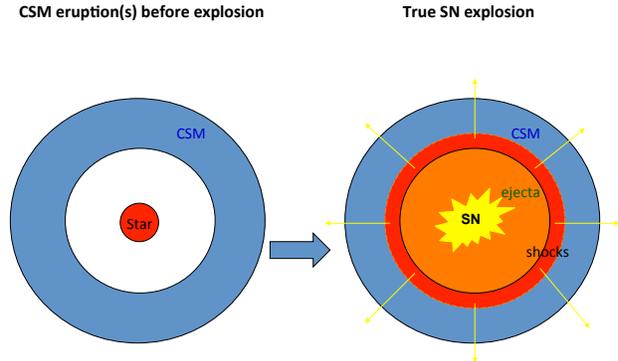}
\caption{The schematic picture of the interaction-powered SN scenario. 
}
\end{figure}

\section[]{Basic Setup}
In this preparatory section, before we discuss non-thermal signatures, we explain the picture of interaction-powered SNe and describe the basic physical setup.

Let us consider SN ejecta with the kinetic energy ${\mathcal E}_{\rm ej}$ and the velocity $V_{\rm ej}$.  Noting ${\mathcal E}_{\rm ej}=M_{\rm ej}V_{\rm ej}^2/2$ for the ejecta mass $M_{\rm ej}$, the momentum and energy conservation laws give
\begin{eqnarray}
M_{\rm ej}V_{\rm ej}+M_{\rm cs}V_{\rm cs}&=&(M_{\rm ej}+M_{\rm cs})V\\
\frac{1}{2}M_{\rm ej}V_{\rm ej}^2+\frac{1}{2}M_{\rm cs}V_{\rm cs}^2&=&\frac{1}{2}(M_{\rm ej}+M_{\rm cs})V^2+{\mathcal E}_d,
\end{eqnarray}
where $M_{\rm cs}$ is the total CSM mass and $V_{\rm cs}(<V_{\rm ej})$ is the CSM velocity.  The total dissipated energy ${\mathcal E}_d$ is written as 
\begin{eqnarray}
{\mathcal E}_d&=&\frac{M_{\rm cs}}{M_{\rm ej}+M_{\rm cs}}\frac{1}{2}M_{\rm ej}{(V_{\rm ej}-V_{\rm cs})}^2\nonumber\\
&\approx&\frac{M_{\rm cs}}{M_{\rm ej}+M_{\rm cs}}{\mathcal E}_{\rm ej},
\end{eqnarray}
where $V_{\rm ej}\gg V_{\rm cs}$ is used in the last equality.  The above equation suggests that a significant fraction of ${\mathcal E}_{\rm ej}$ can be dissipated if the CSM mass is large~\cite[see also, e.g.,][]{van+10,mor+13b}.  Density profiles of both the ejecta and CSM are important for detailed predictions.  For example, when the density profile of the ejecta is steep enough and most of its energy is carried by lower-velocity ejecta material, the explosion has driven waves that can be described by Chevalier-Nadezhin self-similar solutions~\cite{che82}.  When the shock wave sweeps up ambient mass comparable to $M_{\rm ej}$ and it is non-radiative, we expect blast waves that can be described by Sedov-Taylor-like self-similar solutions~\citep[see][and references therein]{tm99}.  
In this work, to push the basic idea and avoid uncertainty in the ejecta profile and many other complications due to radiation processes, we discuss non-thermal properties without relying on such details.  Our treatment still provides an order of magnitude estimate of expected non-thermal signals, and a more detailed study will be presented in an accompanying paper~\cite{mur+14}.  

Hereafter, we assume that the CSM has a wind-like power-law density profile and extends to the edge radius of the wind, $R_w$.  We expect that this is reasonable~\citep[see, e.g.,][]{ofe+13d}, although details are uncertain due to poor understandings of the CSM eruption mechanism.  Then, the CSM density is written as
\begin{equation}
\varrho_{\rm cs}=DR_0^{-2}{\left(\frac{R}{R_0}\right)}^{-s}\simeq5.0\times{10}^{16}~D_{*}R_0^{-2}{\left(\frac{R}{R_0}\right)}^{-s}~{\rm g}~{\rm cm}^{-3}
\end{equation}
where $R$ should be expressed in cm, $R_0={10}^{15}$~cm, and $D_{*}$ is defined~\footnote{Another definition is $\rho_{\rm cs}=D_*R^{-2}$ that is different from ours.} for the mass-loss rate of $\dot{M}_{\rm cs}\equiv1~M_{\odot}~{\rm yr}^{-1}~\dot{M}_{\rm cs,0}$ and the wind velocity of $V_{\rm cs}\equiv{10}^3~{\rm km}~{\rm s}^{-1}~(V_{\rm cs}/{10}^{3}~{\rm km}~{\rm s}^{-1})$.  This can also be expressed by
\begin{equation}
D\equiv\frac{\dot{M}_{\rm cs}}{4\pi V_{\rm cs}}\simeq5.0\times{10}^{16}~\dot{M}_{\rm cs,0}{(V_{\rm cs}/{10}^{3}~{\rm km}~{\rm s}^{-1})}^{-1}~{\rm g}~{\rm cm}^{-1}.
\end{equation}
The CSM mass within $R$ is estimated to be
\begin{equation}
M_{\rm cs}(<R)=\int_{R_{\rm cs}}^{R}dr\,\,\,4\pi r^2\varrho_{\rm cs},
\end{equation}
where $R_{\rm cs}$ is the CSM inner edge radius.  In particular, in the wind case ($s=2$), we have
\begin{equation}
M_{\rm cs}(<R)=4\pi D{\Delta R}\simeq3.2~M_{\odot}D_*R_{16},  
\end{equation} 
where we have used ${\Delta R}\approx R$ and $R\equiv{10}^{16}~{\rm cm}~R_{16}$.  Note that, in the one-zone model where the calculation is performed for a CSM density $n_{\rm cs}$ at a given radius $R$, qualitative pictures for different density profiles are simply obtained by using $M_{\rm cs}$ instead of $D_*$ (although the dynamics and temporal evolution depend on density profiles).
The deceleration is significant after the ejecta accumulates the CSM mass equivalent to its own mass, whose radius is characterized by
\begin{equation}
R_{\rm dec}\approx\frac{M_{\rm ej}}{4\pi D}\simeq{10}^{16}~{\rm cm}~(M_{\rm ej}/{10}^{0.5}~M_{\odot})D_{*}^{-1}. 
\end{equation}
If $R_{\rm dec}<R_w$, most of the ejecta energy is dissipated by the ejecta-CSM collision. 

One of the important quantities is the Thomson optical depth.  Using the CSM electron density,
\begin{equation}
n_e=\frac{DR^{-2}}{\mu_em_H}\simeq3.0\times{10}^{8}~{\rm cm}^{-3}~\mu_e^{-1}D_{*}R_{16}^{-2},
\end{equation} 
the Thomson optical depth is estimated to be
\begin{equation}
\tau_T^u=\int_R dr\,\,\,n_{e}\sigma_T \approx n_e\sigma_T R\simeq2.0~\mu_e^{-1}D_*R_{16}^{-1},
\end{equation}  
for $R<R_w$, where $\sigma_T$ is the Thomson cross section.  The Thomson optical depth in the downstream is also $\tau_T\approx n_e\sigma_T R$ although the density in the thin, interacting shell is compressed by the shock compression ratio.  As seen below, the emission is mostly observed when $\tau_T\lesssim c/V_s$ after photons can leave the system.  However, while the interaction with a dense CSM shell happens at $\tau_T\gtrsim{\rm a~few}$, hard X rays and soft gamma rays produced at the shock cannot avoid Compton down-scattering and a significant fraction of the emission would be thermalized~\cite{ci12,svi+12,pan+13}.  Equal lines of various optical depths, CSM density and luminosities in the ($R$, $D$) plane and ($R$, $M_{\rm cs}$) plane are shown in Figures~2 and 3, respectively.  The equal CSM density line, $D_*\simeq33R_{16}^2n_{\rm cs,10}$, is also overlaid.  Explanations for the optical depths other than the Thomson optical depth are given below.  As argued by Murase et al. (2011), efficient CR acceleration is possible at $\tau_T\lesssim c/V_s$, and for $\tau_{pp}\gtrsim1$ we expect almost all the accelerated CR ions to produce neutrinos, hadronic gamma rays, and secondary electrons and positrons.  In the system, hadronic gamma rays can interact with photons via the two-photon annihilation process and/or matter via the Bethe-Heitler (BH) pair-production process, respectively (see Section 3).  The attenuation of GeV gamma rays due to the BH process is insignificant at $\tau_{\rm BH}\lesssim1$, which is not far from $\tau_{T}\lesssim c/V_s$.

In this work, we consider the forward shock, so the shock velocity $V_s$ is regarded as the forward shock velocity $V_f$.  The reverse shock power is smaller when the ejecta profile is steep~\citep[e.g.,][]{cf03}.  But this might not be the case if the profile is changed, e.g., possibly by experiencing many interactions with many CSM shells.

\begin{figure}
\includegraphics[width=\linewidth]{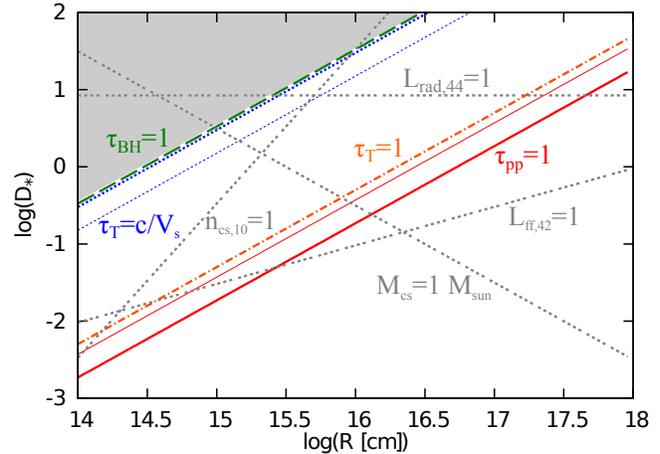}
\caption{The parameter range allowing production and escape of gamma rays in the ($R$, $D$) plane.  See the text for meanings of $\tau_T=c/V_s$, $\tau_T=1$, $\tau_{pp}=1$ and $\tau_{\rm BH}=1$.  The shaded region suggests the range where we do not expect either production or escape of gamma rays.  The thick curves represent $V_s=5000~{\rm km}~{\rm s}^{-1}$ while the thin curves do $V_s={10}^4~{\rm km}~{\rm s}^{-1}$.  With quadruplicate-dotted cures, lines of the constant CSM density ($n_{\rm cs}$=const.), constant post-breakout radiation luminosity ($\epsilon_\gamma L_{\rm kin}$=const.), constant optically-thin free-free luminosity ($L_{\rm ff}$=const.), and constant CSM mass ($M_{\rm cs}$=const.) are also shown for comparison.
}
\end{figure}
\begin{figure}
\includegraphics[width=\linewidth]{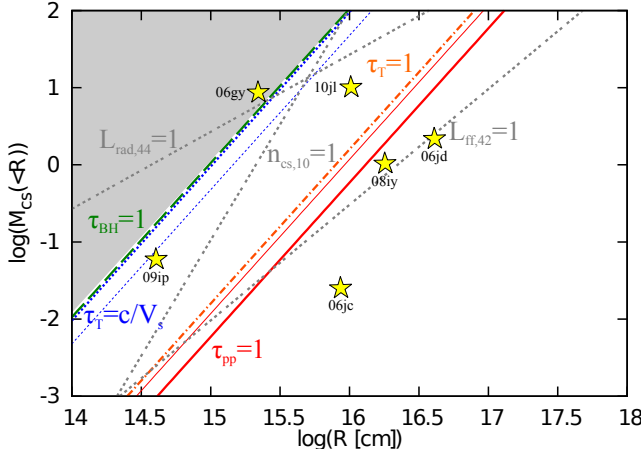}
\caption{The same as Figure~2, but for the ($R$, $M_{\rm cs}$) plane, where $M_{\rm cs}$ has the solar-mass unit.  Lines of the constant CSM density ($n_{\rm cs}$=const.), constant post-breakout radiation luminosity ($\epsilon_\gamma L_{\rm kin}$=const.), and constant optically-thin free-free luminosity ($L_{\rm ff}$=const.) are also shown for comparison.  For explanations of specific SN examples, see Sub-section 2.6.
}
\end{figure}

\subsection{Early phase: subphotospheric interactions}
When a collision with CSM occurs at $\tau_T\gg1$, photons should experience many Compton scatterings, and it takes time for them to leave the system.  The photon diffusion time is roughly $t_{D}\approx{\Delta R}^2\sigma_Tn_{e}/c$, which can further be approximated to be $t_{D}\approx\sigma_T\mu_e^{-1}D_{*}/c$ for the wind profile if $\Delta R\approx R$~\citep[c.f.][]{ci11,bl12}.  Photons cannot essentially diffuse out from the system, when $t_{D}$ is longer than the dynamical timescale is $t_{\rm dyn}\approx R/V_s$.  Hence, when the collision begins at $\tau_T\gtrsim c/V_s$, we start to observe a significant fraction of the emissions in the rise time $t_{\rm rise}$ such that $t_{\rm rise}=t_{D}=t_{\rm dyn}$~\citep[e.g.,][]{sm07,ofe+10,ci11,bl12}.  We define the breakout radius $R_{\rm bo}$, where photons can essentially leave the system.  When the effective diffusion radius $R_D$ is sufficiently smaller than $R_w$ (for $s\geq2$), we have $R_{\rm bo}\approx R_D$, and the shock breakout radius can be written as $R_{\rm bo}\approx V_st_{\rm rise}$.  When the CSM is so dense that $R_D$ is larger than $R_w$, the rise time can be smaller than $R_{\rm bo}/V_s$ because $\Delta R\ll R_{\rm bo}\approx R_w$ at the breakout~\cite{ci11}.

We expect that thermal radiation carries a significant fraction ($\epsilon_\gamma<1$) of the energy carried by the interacting shell ($\mathcal E$), where the radiated energy~\footnote{For the adiabatic index $\hat{\gamma}=4/3$, $\epsilon_\gamma=0.32$ is obtained in the mini-shell model~\cite{ci11}.  See also Ofek et al. (2014).} is ${\mathcal E}_{\rm rad}\equiv\epsilon_\gamma{\mathcal E}$ and $\mathcal E$ after the collision is roughly comparable to ${\mathcal E}_d$ (see Equation~3).  Noting ${\mathcal E}/t_{\rm dyn}\sim L_{\rm kin}\approx(1/2)\varrho_{\rm cs}V_s^3(4\pi R^2)$, the (bolometric) radiation luminosity just after the breakout is  
\begin{eqnarray}
L_{\rm rad}=\epsilon_\gamma L_{\rm kin}&=&\epsilon_{\gamma}\frac{1}{2}\varrho_{\rm cs}V_s^3(4\pi R^2)\nonumber\\
&\simeq&1.3\times{10}^{43}~{\rm erg}~{\rm s}^{-1}(\epsilon_\gamma/0.3)\nonumber\\
&\times&D_{*}~{(V_s/5000~{\rm km}~{\rm s}^{-1})}^3.
\end{eqnarray}
The constant radiation luminosity line, $D_*\simeq8.5{(V_s/5000~{\rm km}~{\rm s}^{-1})}^{-3}L_{\rm rad,44}$ (with $\epsilon_\gamma=0.3$), is depicted in Figures~2 and 3.  A more sophisticated model is given by Chevalier \& Irwin (2011), which is summarized in Appendix A~\footnote{For the purpose of modeling observed light curves, which is not the focus of this work, one may use expressions for more general profiles of $\varrho_{\rm cs}\propto R^{-s}$ and $\varrho_{\rm ej}\propto R^{-m}$~\cite{svi+12,ofe+13d}}.

\subsection{Late phase: post-breakout interactions}
The collision between the SN ejecta and the CSM may start from the optically-thick regime.  Then, after the shock breakout, the interaction eventually enters the optically-thin regime.  This regime typically comes after the time of $\sim(c/V_s)(R_{\rm bo}/V_s)$ when $R_w$ is large enough.  Hence, for optically SLSNe such as SN 2006gy, it usually happens only after the shock crosses $\sim R_w$.  In order to expect optically-thin ejecta-CSM interactions within $R_w$, relatively large $R_{w}$ and/or low $M_{\rm cs}$ are needed.  Alternately, the collision may occur at $\tau_T\lesssim c/V_s$ if CSM effectively has an inner edge and can be regarded as a shell.  Especially for optically-thin interactions at $\tau_T\lesssim1$, hard X rays easily leave the system although ultraviolet photons and soft X rays may be attenuated due to bound-free absorption.  Indeed, such X-ray and radio emissions have been observed in some SNe like SN 1988Z~\cite{cd94,ofe+13b} and 2006jd~\cite{cha+12}. 

Let us consider a CSM extending to $R_w$.  For $\tau_T(R_w)\lesssim c/V_s$, the characteristic duration of emission is expected to be $\approx R_w/V_s$ (or $\approx R_{\rm dec}/V_s$).  When the dominant loss process for thermal electrons is the free-free emission, the (bolometric) radiation luminosity mainly comes from bremsstrahlung emission.  In the non-radiative case, we have~\cite{rl86,cf03}
\begin{eqnarray}
L_{\rm rad}\approx L_{\rm ff}&=&4\pi\Lambda_{\rm ff}n_e^2R^2\approx\Lambda_{\rm ff}\frac{M_{\rm cs}\sigma\varrho_{\rm cs}}{\mu_e^2m_H^2}\nonumber\\
&\simeq&1.1\times{10}^{42}~{\rm erg}~{\rm s}^{-1}~\bar{g}_{\rm ff}\mu_e^{-2}\nonumber\\
&\times&D_{*,-1}^2R_{16}^{-1}(V_s/5000~{\rm km}~{\rm s}^{-1}),
\end{eqnarray}
where $\bar{g}_{\rm ff}$ is the Gaunt factor, $\Lambda_{\rm ff}$ for hydrogen is used, and absorption is not considered yet~\footnote{Even if the absorption is serious, X rays are detectable if the unattenuated X-ray luminosity is large enough, as suggested in SN 2010jl~\cite{ofe+13d}.}.  Here the compression factor $\sigma$ is taken to be 4 but can be larger.  The constant radiation luminosity line, $D_*\simeq0.096\bar{g}_{\rm ff}^{-1/2}\mu_eR_{16}^{1/2}{(V_s/5000~{\rm km}~{\rm s}^{-1})}^{-1/2}L_{\rm ff,42}$, is depicted in Figures~2 and 3.  When the shock is radiative, $L_{\rm rad}$ should be limited by the kinetic luminosity as in equation~(11).  It roughly occurs when the cooling time $t_{\rm ff}\approx(3kT_e/\Lambda_{\rm ff}n_{\rm cs})\simeq3.1~{\rm yr}~\bar{g}_{\rm ff}^{-1}T_{e,8}^{1/2}D_{*,-1}R_{16}^{2}$ is shorter than time $t$.  Note that $L_{\rm ff}$ is proportional to $t^{-1}$ and $T_e\gtrsim2\times{10}^7$~K is assumed, otherwise cooling by line emissions becomes relevant~\citep[e.g.,][]{cf03}.

\subsection{Shock properties}
Whether efficient CR acceleration occurs or not depends on the shock properties. 
When an ejecta-CSM collision occurs at sufficiently small radii, the shock is initially radiation-mediated in the sense that the upstream shock structure is modified by radiation from the downstream.  For non-relativistic shocks where effects of pairs are irrelevant, the shock is radiation-mediated when $\tau_T\gtrsim c/V_s$~\cite{wea76,kat+10}.  When coupling with radiation is strong enough, electrons transfer their energy to photons.  In the thermal equilibrium limit, at the far downstream, photons, electrons and protons have the temperature of
\begin{equation}
kT_{\rm ph}\simeq1.2~{\rm eV}~\tilde{\epsilon}_{\rm gb}^{-1/4}D_*^{1/4}R_{16}^{-1/2}{(V_s/5000~{\rm km}~{\rm s}^{-1})}^{1/2}, 
\end{equation}
where $\tilde{\epsilon}_{\rm gb}aT_{\rm ph}^4=(18/7)\varrho_{\rm cs}V_s^2$ is used and $\tilde{\epsilon}_{\rm gb}$ is the gray-body factor.  Note that this emission does not have to be the observed emission since photons start to escape only after $\tau_T\sim c/V_s$.  Thermal equilibrium may not be realized if sufficient photons are not produced by the bremsstrahlung process~\cite{ci11,svi+12}.  At the nearer downstream, protons and electrons have much higher temperatures. In the absence of collisionless-plasma processes, the electron temperature is determined by the balance between Coulomb heating and cooling processes.  When the relevant cooling process is the Compton cooling, we have~\cite{wl01,mur+11,kat+11}
\begin{equation}
kT_e\sim40~{\rm keV}~\tilde{\epsilon}_{\gamma}^{-2/5},
\end{equation}
over the length scale of $V_s \nu_{ie}^{-1}$ (where $\nu_{ie}$ is the ion-electron collision frequency), where~\footnote{Note that $\tilde{\epsilon}$ is defined for the downstream energy density.  On the other hand, as in Murase et al. (2011), $\epsilon$ is defined for the total energy while $\varepsilon$ is defined for the ram pressure of the upstream flow.} $\tilde{\epsilon}_\gamma=2U_\gamma/(3\sigma n_{\rm cs}kT_p)$, where $U_\gamma$ is the energy density of photons.  Note that the above equation is valid for sufficiently high-velocity shocks, since $T_e$ is limited by the proton temperature $T_p$~\cite{ci12}.  In reality, collisionless-plasma processes can be faster than Coulomb collisional processes, where $T_e$ can be higher than in equation~(14), but should be lower than the equipartition temperature.   

After $\tau_T\lesssim c/V_s$, the shock is no longer radiation-mediated, and we expect collisionless (or collisional) shocks.  For strong, non-relativistic shocks, the proton temperature at the immediate downstream is 
\begin{equation}
kT_p=\frac{2(\hat{\gamma}-1)}{{(\hat{\gamma}+1)}^2}m_pV_s^2,
\end{equation}
where $\hat{\gamma}$ is the adiabatic index.  When the adiabatic index is $\hat{\gamma}=5/3$, we have a well-known result, $T_p=(3/16)m_pV_s^2$.  
The electron temperature is affected by energy transfer from protons, which may be Coulomb heating or faster collisionless plasma processes.  If $\hat{\gamma}=5/3$ and when electrons and protons achieve the equipartition, we have~\citep[e.g.,][]{fra+96,cf03,ofe+13d}
\begin{equation}
kT_{e}\simeq24~{\rm keV}~{(V_s/5000~{\rm km}~{\rm s}^{-1})}^2. 
\end{equation}

Therefore, in the interaction-powered SN scenario, we can naturally expect X rays via bremsstrahlung or inverse-Compton (IC) processes since electrons should be heated by the shock~\citep[e.g.,][]{mur+11,kat+11,ci12,pan+13}.  Also, we may assume the material is highly ionized at least in the immediate upstream.  Detections of X rays allow us to probe the existence of strong shocks, supporting the scenario~\cite{kat+11,ofe+13b}.  However, there are several complications.  First, free-free emission may be suppressed if thermal electrons mainly cool via the IC process.  Secondly, when the ejecta-CSM interaction occurs at $\tau_T\gtrsim1$, hard X rays lose their energies in both the emission zone (downstream) and screen zone (upstream), and softer X rays are down-graded via bound-free absorption.  If the amount of non-ionized atoms similarly exists in the far upstream, the bound-free optical depth for soft X rays is roughly estimated to be~\citep[e.g.,][]{rl86,ofe+13b}
\begin{equation}
\tau_{\rm bf}^u\simeq600~D_*R_{16}^{-1}{(h\nu/\rm keV)}^{-2.5},
\end{equation}
at $h\nu\sim0.03-10$~keV.  Naively, the X-ray luminosity is then~\citep[e.g.,][]{fra+96,cf03}
\begin{equation}
L_X\sim f_{\rm sup}L_{\rm rad}\frac{(1-e^{-\tau})}{\tau}e^{-\tau^u},
\end{equation}
where $f_{\rm sup}$ is the suppression by other losses, $\tau\sim\tau_T$ is the optical depth in the emission zone and $\tau^u\sim\tau_T^u+\tau_{\rm bf}^u$ is the optical depth in the screen zone.  Predictions for both the thermal and non-thermal X rays depend on details including the frequency-dependent opacity and the ionization in the upstream~\citep[see][and references therein]{ci12,svi+12,pan+13}.  This work does not study X-ray emissions in detail, since non-thermal X rays can be largely contaminated or masked by thermal X rays.

\subsection{Particle acceleration}
Now, we consider particle acceleration.  CR acceleration may become efficient when the shock is no longer radiation-mediated and becomes collisionless, which can be realized when $\tau_T\lesssim c/V_s$~\cite{mur+11,kat+11,kas+13}.  Strong small-scale magnetic fields are expected as a result from plasma instabilities, and MHD mechanisms such as the turbulent dynamo~\footnote{For example, if SNe occur in superbubbles formed by multiple SNe, it is possible to expect turbulent magnetic fields driven by interactions with the inhomogeneous interstellar medium.  Here, the CSM could also be highly turbulent and magnetized before the SN ejecta crashes, since the transiently erupted CSM would also form a shock via interactions with the interstellar medium.  In addition, some observations have suggested that the CSM may be clumpy~\citep[e.g.,][]{cd94,smi+09}, and the shocked CSM could achieve strong magnetic fields via the turbulent dynamo due to the ejecta-CSM interaction.} can also play crucial roles especially in the downstream~\citep[e.g.,][]{ino+09,ino+12}.  In addition, CRs themselves excite turbulence via CR stream instabilities, which can be important in the upstream~\citep[see][and references therein]{bel04,bel13,byk+13}.  Since details of these processes are uncertain, for simplicity, we parameterize the magnetic field with the ratio of $B^2/8\pi$ to $\varrho_{\rm cs}V_s^2/2$ as $\varepsilon_{B}\equiv B^2/(4\pi\varrho_{\rm cs}V_s^2)$.  Then, we obtain 
\begin{equation}
B\simeq4.0~{\rm G}~\varepsilon_{B,-2}^{1/2}D_{*}^{1/2}R_{16}^{-1}{(V_s/5000~{\rm km}~{\rm s}^{-1})}.
\end{equation}
In the Bohm limit, the proton acceleration time scale is estimated to be~\citep[see a review][]{dru83} 
\begin{eqnarray}
t_{p-\rm acc}&=&\frac{20}{3}\frac{c^2}{V_s^2}\frac{E_p}{eBc}\\
&\simeq&0.67~{\rm s}~\varepsilon_{B,-2}^{-1/2}D_{*}^{-1/2} R_{16}\nonumber\\
&\times&{(V_s/5000~{\rm km}~{\rm s}^{-1})}^{-3}(E_{p}/{\rm GeV})\nonumber,
\end{eqnarray}
and the proton maximum energy is estimated by comparing it to various cooling time scales (see below).  In the fully-ionized plasma (that is the case in the vicinity of the shock), the Coulomb cooling time of thermal protons~\cite{sch02},
\begin{eqnarray}
t_{p-\rm C}&\approx&\frac{(m_p c^2/{\rm GeV})}{3.1\times{10}^{-16}n_e}\frac{8.3\times{10}^{-9}T_{e,4}^{3/2}+\beta_p^3}{2}\nonumber\\
&\sim&2.4\times{10}^1~{\rm s}~\mu_e^{-1}D_*^{-1}R_{16}^2{(V_s/5000~{\rm km}~{\rm s}^{-1})}^3,
\end{eqnarray}
where $\beta_p\sim V_s/c$ is assumed but the velocity of injected protons may be higher, depending on details of injection processes.  The ratio between the two is $t_{p-{\rm acc}}/t_{p-{\rm C}}\sim2.8\times{10}^{-2}\mu_e\varepsilon_{B,-2}^{-1/2}D_*^{1/2}R_{16}^{-1}{(V_s/5000~{\rm km}~{\rm s}^{-1})}^{-6}(E_p/{\rm GeV})$ when the upstream is cold enough.  Hence, the Coulomb energy loss timescale is longer than their acceleration time, so CR proton acceleration is possible.  Note that the Coulomb cooling is much less relevant in the immediate downstream, where the temperature is much higher.  

As protons are accelerated above the pion production threshold, $\gamma_p>1.37$, inelastic $pp$ interactions occur, leading to production of electrons and positrons via $\pi^\pm\rightarrow\nu_\mu+\bar{\nu}_{\mu}+\nu_e(\bar{\nu}_e)+e^\pm$.  Their minimum injection Lorentz factor is~\cite{der86}
\begin{equation}
\gamma_h\approx\frac{1}{4}\frac{m_\pi}{m_e}\sim68.
\end{equation}
The important point here is that the minimum injection Lorentz factor is unique for hadronic injections, which is different from the case of primary electron acceleration.

Primary electrons can also be shock accelerated via the Fermi acceleration mechanism.  However, the Larmor radius of thermal electrons is smaller than that of thermal protons.  Thus, electrons need to be energized via some plasma processes to cross the shock length and to get injected to the Fermi acceleration process.  In other words, the Larmor radius of relativistic electrons, $\gamma_e m_e c^2/(eB)$, should be larger than that of thermal protons, $\sim cm_pV_s/(eB)$.  The Lorentz factor of elections that can be accelerated by the conventional shock acceleration mechanism satisfies
\begin{equation}
\gamma_{e}\gtrsim\frac{m_p}{m_e}\frac{V_s}{c}\simeq31~(V_s/5000~{\rm km}~{\rm s}^{-1}).
\end{equation}
Keeping this in mind and introducing the energy fraction ($\epsilon_e$) and number fraction ($f_e$) of relativistic electrons distributed with a power law, the injection Lorentz factor of primary shock-accelerated electrons ($\gamma_l$) is expressed as 
\begin{eqnarray}
\gamma_{l}&\approx&g_{q_e}\frac{\epsilon_e}{f_e}\frac{m_p}{m_e}\frac{V_s^2}{2c^2}\nonumber\\
&\simeq&5.1~\epsilon_{e,-3}f_{e,-5}^{-1}(g_{q_e}/0.2){(V_s/5000~{\rm km}~{\rm s}^{-1})}^2, 
\end{eqnarray}
where $g_{q_e}=1/\ln(\gamma_{e}^M/\gamma_{l})$ for $q_e=2$ and $g_{q_e}=(q_e-2)/(q_e-1)$ for $q_e>2$.  Here, $q_e$ is the injection spectral index of accelerated electrons and $\gamma_e^M$ is the maximum Lorentz factor of accelerated electrons.  Note that equation (24) is obtained from $\int d\gamma_e (dN_e/d\gamma_e)=f_e N_p$ and $\int d\gamma_e (\gamma_e m_ec^2)(dN_e/d\gamma_e)=\epsilon_e N_p m_p V_s^2/2$.  The values of $\epsilon_e$ and $f_e$ are uncertain.  The CR spectra observed at the Earth imply $\epsilon_e\sim{10}^{-3}-{10}^{-2}$, and such values are inferred from modeling of radio emission from Type IIb SNe~\citep[e.g.,][]{mae12}.  Smaller values of $\epsilon_e\sim{10}^{-4}$ are obtained in the leptonic scenario for young SN remnants~\citep[e.g.,][]{kw08}.  Sufficiently large values of $f_e$ imply $\gamma_l\lesssim(m_p/m_e)(V_s/c)$, where other electron acceleration processes in the shock transition layer should be relevant, and the spectrum for $\gamma_l\lesssim\gamma_e\lesssim(m_p/m_e)(V_s/c)$ may be steeper than that for $\gamma_e\gtrsim(m_p/m_e)(V_s/c)$.      

The energy carried by accelerated CRs can be summarized as follows.  The CR acceleration is efficient only after the radiation escapes from the system and the strong shock jump is formed by collisionless shocks.  Hence, if ${\rm min}[R_{\rm dec},R_w]<R_{\rm bo}$, we do not expect many CRs.  Normal SNe correspond to $R_{\rm bo}<R_w<R_{\rm dec}$, so only a fraction of the SN explosion energy ${\mathcal E}_{\rm ej}$ is converted to CRs at the time the shock reaches $R_w$.  If the CSM is massive and $R_{\rm bo}<R_{\rm dec}<R_w$ (thus $\tau_T(R_{\rm dec})<c/V_s$), we expect ${\mathcal E}_d\approx{\mathcal E}_{\rm ej}$, so a significant fraction of the SN explosion energy can be converted into the energy of CRs. 
The energy of accelerated CR protons in interaction-powered SNe (${\mathcal E}_{{\rm CR}p}^{\rm ipsn}$) is roughly estimated to be  
\begin{equation}
{\mathcal E}_{{\rm CR}p}^{\rm ipsn}=
\left\{ \begin{array}{ll} 
\epsilon_p{\mathcal E}_d\approx\epsilon_p{\mathcal E}_{\rm ej},
& \mbox{($R_{\rm bo}<R_{\rm dec}<R_w$)}\\
\epsilon_p{\mathcal E}_d\approx\epsilon_p(M_{\rm cs}/M_{\rm ej}){\mathcal E}_{\rm ej}
& \mbox{($R_{\rm bo}<R_w<R_{\rm dec}$)}\\
\ll{\mathcal E}_{\rm ej}
& \mbox{(${\rm min}[R_{\rm dec},R_w]<R_{\rm bo}$)}
\end{array} \right.
\end{equation}
Here equation (3) is used.  Note that, for $R_w<R_{\rm dec}$ (i.e., $M_{\rm cs}<M_{\rm ej}$), only the fraction of the SN explosion energy can be dissipated by one collision.  Also, $\epsilon_p$ is the energy fraction carried by CRs above $m_pc^2$, and $\epsilon_p\sim0.1-0.3$ is typically used in the context of SN remnants. 

\subsection{Fate of cosmic rays: hadronuclear reactions}
When particles are accelerated up to very high energies, high-energy gamma rays and neutrinos should be accompanied via hadronuclear interactions like the $pp$ reaction.  In particular, CR protons interact with nucleons while they are advected to the downstream, so neutrinos and pionic gamma rays are expected in the GeV-PeV range~\cite{mur+11}.  By comparing the $pp$ interaction time scale ${(n_N\sigma_{pp}c)}^{-1}$ and $t_{\rm dyn}$, we get the $pp$ optical depth as  
\begin{equation}
\tau_{pp}\approx\sigma_{pp}n_{\rm cs}R(c/V_s)\simeq5.4~D_{*}R_{16}^{-1}{(V_s/5000~{\rm km}~{\rm s}^{-1})}^{-1},
\end{equation}
where $\sigma_{pp}\approx3\times{10}^{-26}~{\rm cm}^2$ is used.  From Figures~2 and 3, one sees large parameter space satisfying $\tau_T\lesssim c/V_s$ and $\tau_{pp}\gtrsim1$, where neutrinos and gamma rays provide promising signals of CR proton acceleration at collisionless shocks.  Even if $\tau_{pp}\lesssim0.1$, we can say that a significant fraction of the CR energy is converted into hadronic emission via $pp$ interactions.  From equation (26), one sees $D_*\propto \tau_{pp}RV_s$ or $M_{\rm cs}(<R)\propto \tau_{pp}R^2V_s$.  Thus, as seen from Figure~3, $pp$ processes are typically efficient for ejecta-CSM interactions at radii of $R\lesssim{10}^{16.5}$~cm, for $M_{\rm cs}\sim1M_{\odot}$.  Note that $pp$ interactions are relevant even in the optically-thin regime of $\tau_T\lesssim1$.

\subsection{Examples}
We here discuss several examples of SNe to see if ejecta-CSM interactions satisfying $\tau_T\lesssim c/V_s$ and $\tau_{pp}\gtrsim0.1$ are indeed indicated by recent observations.

First, we consider the optically-thick regime around the shock breakout.  Observationally, the CSM nucleon density $n_{\rm cs}$ (or $D_*$) can be estimated from the radiation luminosity $L_{\rm rad}$ (or the radiated energy ${\mathcal E}_{\rm rad}$) at the time of the shock breakout, the rise time $t_{\rm rise}$, and the shock velocity $V_s$ (or $R_{\rm bo}$ that is the radius at the shock breakout).  Note that the approximation $\Delta R\approx R$ is valid when $R_w>R_{D}=\sigma_T\mu_e^{-1}D_{*}V_s/c$.  
 
\begin{enumerate}
\item[$\bullet$]SN 2006gy: SN 2006gy is one of the SLSNe~\citep[e.g.,][]{smi+07,ofe+07,smi+10}.  The radiated energy is ${\mathcal E}_{\rm rad}(t_{\rm bo})\approx{10}^{51}~{\rm erg}$ and the breakout time is $t_{\rm rise}\approx60$~d~\cite{sm07,ci11}.  The breakout radius is estimated to be $R_{\rm bo}\approx2\times{10}^{15}$~cm, corresponding to $V_s\sim4000~{\rm km}~{\rm s}^{-1}$.  These observational parameters imply $D_*\sim10$ (see Appendix A), leading to $n_{\rm cs}\sim{10}^{11}~{\rm cm}^{-3}$ and $M_{\rm cs}\sim8M_\odot$ within $R_{\rm bo}$.  These are roughly consistent with numerical modeling of optical light curves~\cite{mor+13}. 
\item[$\bullet$]SN 2009ip: SN 2009ip is one of the luminous Type IIn SNe, which showed re-brightening in 2012.  The radiated energy is ${\mathcal E}_{\rm rad}(t_{\rm bo})\approx1.3\times{10}^{49}~{\rm erg}$ and the rising time~\footnote{The actual shock breakout time scale will be shorter, by a factor of a few, than the visible-light rise time of the SN light curve, since the bolometric light curve of Type IIn SNe seems to rise faster than the optical light curve.  Therefore, with the exception of SN 2010jl, $t_{\rm rise}$ given here, are regarded as an upper limit.} is $t_{\rm rise}\approx10$~d.  The breakout radius is estimated to be $R_{\rm bo}\approx5\times{10}^{14}$~cm, corresponding to $V_s\sim6000~{\rm km}~{\rm s}^{-1}$.  These observational parameters imply $D_*\sim0.3$, $n_{\rm cs}\sim4\times{10}^{10}~{\rm cm}^{-3}$ and $M_{\rm cs}\sim0.05M_\odot$ within $R_{\rm bo}$~\cite{mar+13}, which are consistent with observational constraints~\cite{ofe+13}. 
\end{enumerate}

In Figure~3, both of the examples roughly lie on the $\tau_T=c/V_s$ lines.  The condition $\tau_{pp}>1$ is satisfied, so neutrinos and gamma rays should be produced in the presence of CR protons.  SN 2006gy almost lies on the constant luminosity line of $\epsilon_\gamma L_{\rm kin}={10}^{44}~{\rm erg}~{\rm s}^{-1}$.  Note that ${\mathcal E}_d\approx{\mathcal E}_{\rm ej}$ can be expected for SN 2006gy but not for SN 2009ip. 

Next, we consider post-breakout emission in the later phase.  For $\tau_T\lesssim c/V_s$, the shock crossing time $t_{s}\approx R_w/V_s$ or the deceleration time $t_{\rm dec}\approx R_{\rm dec}/V_s$ can also be used instead of $t_{\rm rise}$.  Then, one can observationally estimate $n_{\rm cs}$ (or $D_*$) from $L_{\rm rad}$, $t_s$ (or $t_{\rm dec}$), and $V_s$.  Alternatively, if we assume that the CSM is not completely ionized, then the X-ray measurements of the bound-free absorption can provide an estimate (or at least a lower limit) on $N_H\approx n_HR$.  Two examples are given below. 

\begin{enumerate}
\item[$\bullet$]SN 2006jd: SN 2006jd was a Type IIn SN, showing bright X-ray emission with the unabsorbed X-ray luminosity of $\sim3\times{10}^{41}~{\rm erg}~{\rm s}^{-1}$ in the $0.2-10$~keV range~\cite{cha+12}.  Radio emission was detected after $\sim400$~d.  With $V_s=5000~{\rm km}~{\rm s}^{-1}$, we expect $R=4\times{10}^{16}~{\rm cm}$ at $t\sim1000$~d.  Then, with the temperature $kT_e\sim60$~keV, $L_{\rm ff}\sim{10}^{42}~{\rm erg}~{\rm s}^{-1}$ implies $n_{\rm cs}\sim4\times{10}^{6}~{\rm cm}^{-3}$ and $M_{\rm cs}\sim2M_\odot$ within $R$.  This estimated CSM density is comparable to $n_{\rm cs}\sim3\times{10}^{6}~{\rm cm}^{-3}$ obtained with $s=1.6$, $\dot{M}_{\rm cs,-3}/V_{\rm cs,2}\approx3.3$ (for $R_0={10}^{15}$~cm), and $R=4\times{10}^{16}$~cm~\cite{cha+12}.  On the other hand, X-ray absorption allows us to estimate the column density of non-ionized hydrogen atoms ($N_H$) that may exist in the far upstream.  It is also suggested that this density is different from values based on X-ray observations, which may imply that the CSM is highly ionized even in the far upstream so there are not many non-ionized atoms in the line of sight of X rays.  However, detections of radio emission may imply the absence of too strong free-free absorption in the upstream.
\item[$\bullet$]SN 2010jl: SN 2010jl was also a Type IIn SNe, showing bright X-ray emission without radio detections.  The unabsorbed X-ray luminosity of $\sim7\times{10}^{41}~{\rm erg}~{\rm s}^{-1}$ is obtained in the $0.2-10$~keV range~\cite{cha+12b}.  However, Ofek et al. (2014) suggested a higher intrinsic luminosity and obtained $N_H\sim{10}^{25}~{\rm cm}^2$ from modeling of the optical emission in the early phase around $t_{\rm bo}$, which is consistent with a value indicated by X-ray observations.  With $V_s=4000~{\rm km}~{\rm s}^{-1}$, we expect $R={10}^{16}~{\rm cm}$ at $t\sim300$~d, implying $n_{\rm cs}\sim{10}^{9}~{\rm cm}^{-3}$ and $M_{\rm cs}\sim10M_\odot$ at that time.  Although there is uncertainty, the CSM with several solar masses is likely~\cite{ofe+13d,fra+13}, and this energetic SN seems one of the promising targets of dedicated searches for neutrinos and gamma rays. 
\end{enumerate}

Based on these parameters, SN 2010jl has $\tau_{pp}>1$, indicating efficient neutrino and gamma-ray production.  On the other hand, $\tau_{pp}<1$ is suggested for SN 2006jd, but the $pp$ efficiency is still significant.  SN 2006jd also lies on the constant luminosity line of $L_{\rm ff}={10}^{42}~{\rm erg}~{\rm s}^{-1}$.  Note that another estimate is possible based on detailed modeling of radio SNe~\citep[e.g.,][]{che82b,che84,che98}, although this work does not focus on such more model-dependent studies.  Most of type IIn SNe have not been seen by low-frequency radio observations, but we show that some of them may be detectable at high-frequency radio wavelengths including mm/submm and FIR bands (see below).   

In the Figure~3, we indicate the above examples by stars for the purpose of demonstration.  Note that their parameters have large uncertainty so such plots do not have to be very precise.  Also, $R$ increases as the observation time $t$, so one can ideally draw evolution curves in the ($R$, $D$) plane and ($R$, $M_{\rm cs}$) plane.  In addition to the four SNe, we indicate SN 2006jc and 2008iy, which are also likely to be interaction-powered SNe.  For SN 2006jc, we use $M_{\rm cs}\sim0.02M_\odot$ and $R=9\times{10}^{15}$~cm at the X-ray peak time of $t\sim110$~d based on Immler et al. (2008).  For SN 2008iy, we adopt $M_{\rm cs}\sim1M_\odot$ and $R=1.7\times{10}^{16}$~cm at the peak time of $t\sim400$~d from Miller at al. (2010).

\subsection{Maximum energy: possible Pevatrons}
SN remnants have been largely believed to be the origin of Galactic CRs up to the knee of ${10}^{6.5}$~GeV~\citep[see a review][and references therein]{bel13}.  It would also be interesting to consider interaction-powered SNe as potential accelerators of high-energy CRs.  The maximum energy of accelerated protons, $E_p^M$, is determined by comparing $t_{\rm acc}$ to the cooling time $t_{\rm cool}$ and dynamical time $t_{\rm dyn}$.  

If the time scales of energy losses (including adiabatic losses) are long enough, the maximum energy is limited by the dynamical time, 
\begin{equation}
t_{\rm dyn}\approx\frac{R}{V_s}\simeq2.0\times{10}^{7}~{\rm s}~R_{16}{(V_s/5000~{\rm km}~{\rm s}^{-1})}^{-1}.
\end{equation}
Then, the maximum energy is~\cite{mur+11} 
\begin{equation}
E_{p}^{M}\approx3.0\times{10}^{7}~{\rm GeV}~\varepsilon_{B,-2}^{1/2}D_{*}^{1/2}{(V_s/5000~{\rm km}~{\rm s}^{-1})}^{2}.
\end{equation} 
Note that the Larmor radius ($r_{L}$) of protons is smaller than the system size ($R$), where the protons are confined. 

At small $R$ and/or large $D$, the maximum energy is limited by energy losses.  The $pp$ cooling time of protons is expressed as
\begin{equation}
t_{pp}=\frac{1}{\kappa_{pp}\sigma_{pp}n_{\rm cs}c}\simeq7.4\times{10}^{6}~{\rm s}~D_{*}^{-1}R_{16}^2, 
\end{equation}
where $\kappa_{pp}\approx0.5$ is the proton inelasticity.  Equating $t_{p-\rm acc}$ and $t_{pp}$ gives~\cite{mur+11}
\begin{equation}
E_{p}^{M}\approx1.1\times{10}^{7}~{\rm GeV}~\varepsilon_{B,-2}^{1/2}D_{*}^{-1/2}R_{16}{(V_s/5000~{\rm km}~{\rm s}^{-1})}^{3}.
\end{equation}

Before CRs propagate in a galaxy, they need to escape from the system without significant losses~\citep[e.g.,][]{cap+09,ohi+10,dru11}.  While the ejecta interacts with a CSM, CR escape may be limited by the free escape boundary $l_{\rm esc}$, which could be determined, e.g., by magnetic field amplification processes or wave damping via ion-neutral collisions~\cite{kc71}.  By comparing the diffusion length $(1/3)(c r_L/V_s)$ (in the Bohm limit) to $l_{\rm esc}$, we obtain  
\begin{equation}
E_{p}^{\rm max}\approx3.0\times{10}^{7}~{\rm GeV}~\varepsilon_{B,-2}^{1/2}D_{*}^{1/2}(l_{\rm esc}/R){(V_s/5000~{\rm km}~{\rm s}^{-1})}^{2},
\end{equation}
where $E_p^{\rm max}$ is the maximum energy of escaping protons.  If $E_p^M$ is too low for CRs to escape within $t_{\rm dyn}$, CRs are confined and their escape is non-trivial.  If magnetic fields rapidly decay after the shock crossing time (as often supposed in gamma-ray bursts), the condition can be $t_{\rm esc}\sim t_{\rm dyn}<t_{\rm cool}$, otherwise it depends on diffusion and adiabatic losses.  As a reasonable condition for CRs not to be depleted, we here assume $l_{\rm esc}\sim R$ and $\tau_{pp}<1$.  

In Figures~4 and 5, we show the parameter range allowing $E_p^{\rm max}={10}^{15.5}$~eV protons.  Parameter space allowing for higher $E_p^{\rm max}$ is narrower.  If an ejecta-CSM interaction occurs at $\tau_{pp}\gtrsim1$, CRs are largely depleted.  We need $M_{\rm cs}\gtrsim M_{\rm ej}$ to expect ${\mathcal E}_d\sim{\mathcal E}_{\rm ej}$, and $R\gtrsim{10}^{16}$~cm is typically favored for escaping CRs to avoid significant $pp$ cooling.    

Higher-energy protons can generate pairs via the BH process, which occurs at $E_p h\nu>m_pm_ec^4$, i.e., $E_p>4.8\times{10}^{5}~{\rm GeV}~{(h\nu/1~{\rm eV})}^{-1}$.  Sufficiently high-energy protons may dominantly lose their energies via the BH process, especially in the optically-thick regime.  For $\tau_T\gtrsim1$, the number density of thermalized photons is $n_{\rm ph}\sim\tau_TL_{\rm ph}/(4\pi R^2ckT_{\rm ph})$, where $L_{\rm ph}$ is the luminosity of thermalized (re-processed) photons and $\epsilon_{\rm ph}$ is their energy fraction.  Assuming $L_{\rm ph}=\epsilon_{\rm ph}L_{\rm kin}$, the BH cooling time $t_{p-{\rm BH}}\approx1/(\kappa_{\rm BH}\sigma_{\rm BH}n_{\rm ph}c)$ becomes  
\begin{eqnarray}
t_{p-{\rm BH}}&\simeq&3.4\times{10}^7~{\rm s}~\epsilon_{\rm ph}^{-1}\mu_eD_*^{-2}\nonumber\\
&\times&R_{16}^{3}{(V_s/5000~{\rm km}~{\rm s}^{-1})}^{-3}{(kT_{\rm ph}/1~{\rm eV})},
\end{eqnarray}
where $\kappa_{\rm BH}\sigma_{\rm BH}\approx7.6\times{10}^{-31}~{\rm cm}^2$ is used at $h\bar{\nu}\sim20m_ec^2$ in the proton rest frame~\cite{cho+92}.  This is typically longer than the $pp$ cooling time, so we may mainly consider the $pp$ reaction.  
As indicated in Figures~4 and 5, on the other hand, the optically-thin regime of $\tau_T<1$ is more likely in cases where protons with $E_p\sim{10}^{6.5}$~GeV survive.  X-ray photons interact with TeV protons and the number density of optically-thin bremsstrahlung photons is $n_{X}\sim L_{\rm ff}/(4\pi R^2c kT_e)$, so we have
\begin{eqnarray}
t_{p-{\rm BH}}&\sim&1.2\times{10}^{15}~{\rm s}~\mu_e^{2}\bar{g}_{\rm ff}^{-1}D_{*,-1}^{-2}\nonumber\\
&\times&R_{16}^{3}{(V_s/5000~{\rm km}~{\rm s}^{-1})}^{-1}{(kT_e/50~{\rm keV})},
\end{eqnarray}
which is negligible compared to $t_{pp}$.  In addition, photomeson production may also occur at $E_p>6.5\times{10}^{7}~{\rm GeV}~{(h\nu/1~{\rm eV})}^{-1}$.  Although it seems that the proton energy has to be quite high, some interactions with X-ray photons are possible in the attenuation scale of X rays.   

Finally, we briefly discuss a possible contribution of interaction-powered SNe to the observed Galactic CRs.  Contributions from various types of SNe including ``hypernovae''~\footnote{Hypernovae are often defined as SNe with ${\mathcal E}_{\rm ej}>{10}^{52}$~erg, which are usually broadline Type Ibc SNe.  Note that only a fraction of them are trans-relativistic SNe that show a mildly relativistic component.} have been considered~\citep[e.g.,][]{sve03}.  It is still under debade which astrophysical accelerator is responsible for CRs around the knee, although normal SNe have been commonly believed as a leading candidate.  An issue is how CR protons are accelerated up to the knee, and interaction-powered SNe could have some advantages because of higher densities and possible stronger fields.  The observed CR energy flux at $E_{2}={10}^{6.5}$~GeV is smaller than that at $E_{1}=1$~GeV by $\sim2.8\times{10}^{-5}$.  The contribution of interaction-powered SNe at $E_2$ compared to the contribution of normal SNe at $E_1$ is~\citep[e.g.,][]{bud+08} 
\begin{equation}
\frac{E_{2}^2 \Phi_{\rm CR}}{E_1^2 \Phi_{\rm CR}}\approx\frac{\rho_{\rm ipsn}}{\rho_{\rm sn}}\frac{t_{\rm conf}(E_2)}{t_{\rm conf}(E_1)}\frac{{\mathcal E}_{{\rm CR}p}^{\rm ipsn}}{{\mathcal E}_{{\rm CR}p}^{\rm sn}}\frac{{\mathcal R}_p(E_1)}{{\mathcal R}_p(E_2)},
\end{equation}
where ${\mathcal R}_p$ is the conversion factor from the total energy to the energy spectrum (see Appendix A), $\rho_{\rm ipsn}$ is the rate of interaction-powered SNe, $\rho_{\rm sn}$ is the SN rate, and $t_{\rm conf}$ is the confinement time of Galactic CRs.  Although the confinement time is highly uncertain, if $t_{\rm conf}(E_1)/t_{\rm conf}(E_2)\sim{({10}^{6.5}~{\rm GeV}/1~{\rm GeV})}^{1/2}\sim1800$ and ${\mathcal E}_{{\rm CR}p}^{\rm ipsn}\sim{\mathcal E}_{{\rm CR}p}^{\rm sn}$, interaction-powered SNe could contribute to the observed CR flux around the knee if the rate of interaction-powered SNe is as high as $\rho_{\rm ipsn}\sim0.05\rho_{\rm sn}$. 

\begin{figure}
\includegraphics[width=\linewidth]{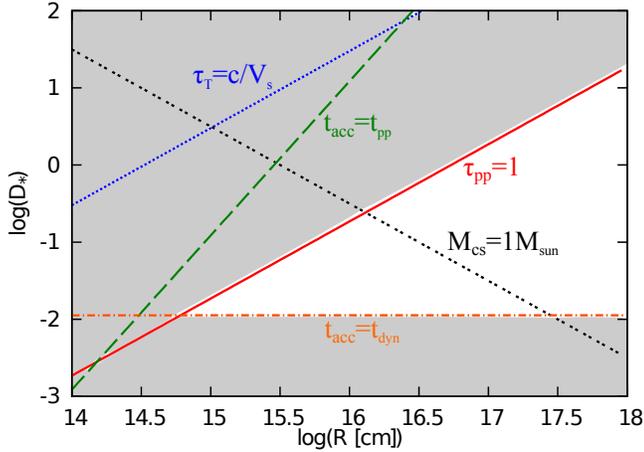}
\caption{The parameter range allowing $E_p^{\rm max}={10}^{15.5}$~eV protons in the ($R$, $D$) plane.  CR acceleration is possible at $\tau_T\lesssim c/V_s$ and $t_{\rm acc}<t_{pp}$ and $t_{\rm acc}<t_{\rm dyn}$ are required to achieve $E_p^M={10}^{15.5}$~eV in the acceleration region.  The shaded region suggests the range where we do not expect either production or escape of ${10}^{15.5}$~eV protons. 
}
\end{figure}
\begin{figure}
\includegraphics[width=\linewidth]{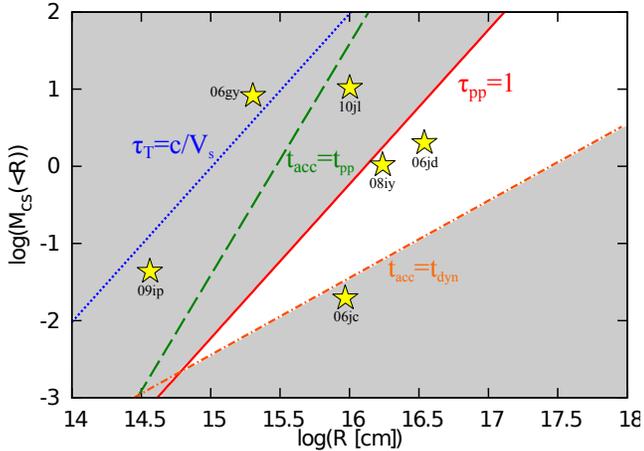}
\caption{The same as Figure~4, but for the ($R$, $M_{\rm cs}$) plane, where $M_{\rm cs}$ has the solar-mass unit.  
}
\end{figure}

\section{High-Energy Neutrino and Gamma-Ray Diagnostics}
By taking into account the inelasticity $\kappa_{pp}\approx0.5$, we obtain the effective $pp$ optical depth as
\begin{equation}
f_{pp}\approx\kappa_{pp}\tau_{pp}\simeq2.7~D_{*}R_{16}^{-1}{(V_s/5000~{\rm km}~{\rm s}^{-1})}^{-1},
\end{equation}
which gives the meson production efficiency.  Using $f_{pp}$, the neutrino energy fluence (per flavor) is estimated to be
\begin{eqnarray}
E_\nu^2 \phi^{\nu}&\approx&\frac{1}{4\pi d^2}\frac{1}{6}{\rm min}[1,f_{pp}]E_p^2\frac{dN_p}{dE_p}\nonumber\\
&\simeq&1.4\times{10}^{-4}~{\rm erg}~{\rm cm}^{-2}~{\rm min}[1,f_{pp}]{\left(\frac{E_\nu}{0.05m_pc^2}\right)}^{2-q_p}\nonumber\\
&\times&\epsilon_{p,-1}{\mathcal E}_{51}{(d/10~{\rm Mpc})}^{-2}{\mathcal R}_{p0,1}^{-1},
\end{eqnarray}
where $q_p$ is the power-law index of the proton spectrum and ${\mathcal R}_{p0}$ is $R_p\equiv\epsilon_p\mathcal{E}/(E_p^2 dN_p/dE_p)$ at $m_pc^2$ (see Appendix A).  When the CSM mass is sufficiently large, assuming $f_{pp}<1$ at the collision, we expect the typical energy fluence of $E_\nu^2 \phi^{\nu}\propto D_{*}R_{\rm dec}^{-1}{V_s}^{-1}$ due to ${\mathcal E}_d\approx {\mathcal E}_{\rm ej}$.  On the other hand, when the CSM mass is not large enough, assuming $f_{pp}<1$ at the collision, the typical energy fluence is expected to be $E_\nu^2 \phi^{\nu}\propto D_{*}^2{V_s}^{-1}$ due to ${\mathcal E}_d\approx(M_{\rm cs}/M_{\rm ej}) {\mathcal E}_{\rm ej}$.  The typical neutrino energy is $E_\nu\sim0.05E_p$, so we expect GeV-PeV neutrino emission given that $E_p^M$ reaches $\sim10-30$~PeV.  The IceCube effective area for muon neutrinos is order of ${10}^{6}~{\rm cm}^2$ in the 100~TeV range while $3\times{10}^{3}~{\rm cm}^2$ in the 1~TeV range, implying the fluence sensitivity of $\sim{10}^{-4}~{\rm erg}~{\rm cm}^{-2}$~\cite{ahr+04}.  Hence, high-energy neutrinos are detectable by IceCube for nearby SNe that occur at $d\lesssim10-20$~Mpc, and stacking analyses for aggregated signals from a number of interaction-powered SNe will also be useful~\cite{mur+11}.  In Appendix A, we provide recipes that connect the observed optical emission to the neutrino signal.
 
The pionic gamma-ray energy fluence is 2 times larger than the neutrino fluence per flavor (after neutrino mixing), so we expect
\begin{eqnarray}
E_\gamma^2 \phi^{\gamma}&\approx&\frac{1}{4\pi d^2}\frac{1}{3}{\rm min}[1,f_{pp}]E_p^2\frac{dN_p}{dE_p}\nonumber\\
&\simeq&2.8\times{10}^{-4}~{\rm erg}~{\rm cm}^{-2}~{\rm min}[1,f_{pp}]{\left(\frac{E_\gamma}{0.1m_pc^2}\right)}^{2-q_p}\nonumber\\
&\times&\epsilon_{p,-1}{\mathcal E}_{51}{(d/10~{\rm Mpc})}^{-2}{\mathcal R}_{p0,1}^{-1},
\end{eqnarray}
or using $E_p^2 dN_p/(dE_pdt)\sim\epsilon_pL_{\rm kin}/{\mathcal R}_p$, the typical gamma-ray energy flux is
\begin{eqnarray}
\nu F_\nu&\approx&\frac{1}{4\pi d^2}\frac{1}{3}{\rm min}[1,f_{pp}]E_p^2\frac{dN_p}{dE_pdt}\nonumber\\
&\sim&1.1\times{10}^{-11}~{\rm erg}~{\rm cm}^{-2}~{\rm s}^{-1}~{\rm min}[1,f_{pp}]\nonumber\\
&\times&{\left(\frac{E_\gamma}{0.1m_pc^2}\right)}^{2-q_p}\epsilon_{p,-1}D_*\nonumber\\
&\times&{(V_s/5000~{\rm km}~{\rm s}^{-1})}^{3}{(d/10~{\rm Mpc})}^{-2}{\mathcal R}_{p0,1}^{-1}.
\end{eqnarray}
The \textit{Fermi} 1~yr sensitivity at GeV energies is $\sim3\times{10}^{-12}~{\rm erg}~{\rm cm}^{-2}~{\rm s}^{-1}$~\cite{fh13}, so interaction-powered SNe may be detectable with \textit{Fermi} for powerful explosions at $d\lesssim20-30$~Mpc~\cite{mur+11}.  Stacking analyses are useful for GeV gamma rays as well as neutrinos (see Appendix A).  In addition, future ground Cherenkov detectors will be more sensitive.  For example, the CTA 100~hr sensitivity at TeV is $\sim{10}^{-13}~{\rm erg}~{\rm cm}^{-2}~{\rm s}^{-1}$~\cite{cta11}, whereas the HAWC 3~yr sensitivity at 2~TeV is $\sim3\times{10}^{-12}~{\rm erg}~{\rm cm}^{-2}~{\rm s}^{-1}$~\cite{abe+13}.  If the CR spectrum is as hard as $q_p\sim2$, we expect that detections are possible up to $d\lesssim100-200$~Mpc via followup observations by CTA within days-to-years.   

For extragalactic gamma rays, one has to keep in mind attenuation processes.  There are two effects: attenuation by the extragalactic background light (EBL) and attenuation by target photons in the source.  The attenuation by EBL is relevant above $\sim10-100$~TeV~\cite{mur+11,kas+13}, so we can neglect it to discuss the detestability in the GeV-TeV range.  The attenuation in the source can be relevant especially for emission from the reverse shock, but we show below that gamma rays from the shocked CSM can typically escape from the source after the shock breakout, $\tau_T\lesssim c/V_s$.

\subsection{Bethe-Heitler pair-creation process}
Gamma rays around MeV energies are downgraded via Compton scattering with electrons in the matter.  The BH pair-creation process occurs at $E_\gamma\geq2m_ec^2$ and becomes dominant over the Compton scattering above the GeV range, where the Compton scattering is reduced by the Klein-Nishina suppression.  At sufficiently high energies, the approximate cross section of the BH process is~\cite{cho+92} 
\begin{equation}
\sigma_{\rm BH}\approx\frac{3\alpha}{8\pi}\sigma_T\left[\frac{28}{9}\ln\left(\frac{2E_\gamma}{m_ec^2}\right)-\frac{218}{27}\right].
\end{equation}
Then, the BH opacity for gamma rays is expressed as 
\begin{equation}
\tau_{\rm BH}\approx\sigma_{\rm BH}n_{e}R=\frac{\sigma_{\rm BH}}{\sigma_{T}}\tau_T\simeq0.03~D_{*}\mu_{e}^{-1}R_{16}^{-1},
\end{equation}
where $\sigma_{\rm BH}$ is evaluated at $E_\gamma=1~\rm GeV$.  The parameter space such that $\tau_{\rm BH}<1$ is shown in Figures~2 and 3.  Interestingly, it is comparable to the parameter space such that $\tau_T<c/V_s$ for typical shock velocities, since gamma-ray attenuation due to the BH process is irrelevant when $V_s$ is higher than $c\sigma_{\rm BH}/\sigma_T\sim4500~{\rm km}~{\rm s}^{-1}$. 

Note that the BH opacity may be even smaller.  Several observations have suggested that CSM is clumpy rather than uniform.  The BH opacity becomes irrelevant when the CSM is anisotropic or so clumpy that most of the CSM mass is concentrated in dense clumps.  For example, if the density enhancement in the clumps is $\delta\varrho/\varrho\sim100$, the BH opacity is changed by $f_{\rm clu}={(\delta\varrho/\varrho)}^{-2/3}\sim0.05$, where the attenuation by the BH process becomes even less relevant.

\subsection{Two-photon annihilation process}
Gamma rays interact with photons via $\gamma+\gamma\rightarrow e^++e^-$.  The interaction typically happens at $E_\gamma h\nu\approx m_e^2c^4$, and the $\gamma \gamma$ opacity for sufficiently high-energy gamma rays is
\begin{equation}
\tau_{\gamma\gamma}\approx\frac{3}{16}\sigma_{T}(\nu n_{\nu})R,
\end{equation}
where $\nu n_{\nu}$ is the photon number density at $\nu$. 

When the collision with CSM occurs at $\tau_T\gtrsim{\rm a~few}$, most of the emissions are thermalized.  Using $\nu n_\nu\sim n_{\rm ph}\sim L_{\rm ph}\tau_{T}/(4\pi R^2c kT_{\rm ph})$, we obtain
\begin{equation}
\tau_{\gamma\gamma}\simeq1.6\times{10}^3~\epsilon_{\rm ph}\mu_e^{-1}D_*^{2}R_{16}^{-2}{(V_s/5000~{\rm km}~{\rm s}^{-1})}^{3}{(kT_{\rm ph}/1~{\rm eV})}^{-1}
\end{equation}
at $E_\gamma\approx260~{\rm GeV}~{(kT_{\rm ph}/1~{\rm eV})}^{-1}$.  Above this energy, interactions mainly happen in the Klein-Nishina regime, so the $\gamma\gamma$ opacity decreases as $\propto\ln[0.47E_\gamma kT_\gamma/(m_e^2c^4)]E_\gamma^{-1}$.  

Hard X rays, which can be produced by bremsstrahlung emission in the downstream, could potentially prohibit lower-energy gamma rays from leaving the emission region because the pair-creation threshold energy is lower for target photons with higher energies.  Using $E_\gamma h\nu\approx m_e^2c^4$ and $h\nu\sim kT_e$, the typical energy of gamma rays interacting with X-ray photons is estimated to be
\begin{equation}
E_\gamma\sim5.2~{\rm MeV}~{(kT_e/50~{\rm keV})}^{-1}.
\end{equation}

As explained below we expect that GeV gamma rays can leave the system without significant attenuation.  
First, we consider an optically-thin collision at $\tau_T\lesssim1$ with $L_{\rm rad}\approx L_{\rm ff}$.  The black-body approximation is invalid, and hard X-ray emission becomes largely visible.  The number density of optically-thin bremsstrahlung photons is $n_{X}\sim L_{\rm ff}/(4\pi R^2c kT_e)$, so we obtain 
\begin{eqnarray}
\tau_{\gamma\gamma}&\sim&4.5\times{10}^{-4}~\mu_e^{-2}\bar{g}_{\rm ff}\nonumber\\
&\times&D_{*,-1}^2R_{16}^{-2}{(V_s/5000~{\rm km}~{\rm s}^{-1})}{(kT_e/50~{\rm keV})}^{-1},
\end{eqnarray}
so gamma rays will be able to leave the emission region for ejecta-CSM interactions at $\tau_T\lesssim1$.  

The radiation luminosity may be limited by the kinetic luminosity, and interactions with X rays can be relevant in the length scale of $\sim{(n_e\sigma_T)}^{-1}$ even in the optically-thick regime of $1\lesssim\tau_T\lesssim c/V_s$.  
Then, we may roughly expect 
\begin{equation}
\tau_{\gamma\gamma}\sim0.016~\epsilon_\gamma D_*R_{16}^{-1}{(V_s/5000~{\rm km}~{\rm s}^{-1})}^{3}{(kT_e/50~{\rm keV})}^{-1},
\end{equation}
and $\tau_{\gamma\gamma}\sim0.49~\epsilon_\gamma\mu_e{(V_s/5000~{\rm km}~{\rm s}^{-1})}^{2}{(kT_e/50~{\rm keV})}^{-1}$ at $R_{\rm bo}$.  Therefore, given $\epsilon_\gamma<1$, GeV gamma rays would be able to escape from the system.  Note that the BH attenuation is also avoidable for sufficiently high shock velocities.

\section[]{High-Frequency Radio Diagnostics}
Electrons and positrons (for which we simply say electrons) can generate synchrotron emission.  The synchrotron cooling time of relativistic electrons is
\begin{eqnarray}
t_{e-\rm syn}&\approx&\frac{6\pi m_e c}{\sigma_T B^2 \gamma_e}\\
&\simeq&4.9\times{10}^6~{\rm s}~\gamma_{e,2}^{-1}\varepsilon_{B,-2}^{-1}\nonumber\\
&\times&D_{*,-1}^{-1}{(V_s/5000~{\rm km}~{\rm s}^{-1})}^{-2}R_{16}^2.\nonumber  
\end{eqnarray}
On the other hand, the IC cooling timescale is
\begin{equation}
t_{e-\rm IC}=\frac{6 m_e c}{\sigma_T B^2 \gamma_e Y_{\rm IC}},
\end{equation}
where $Y_{\rm IC}=Y_{\rm SSC}+Y_{\rm EIC}$ is the Compton $Y$ parameter, $Y_{\rm SSC}$ is the synchrotron self-Compton (SSC) $Y$ parameter, and $Y_{\rm EIC}$ is the external inverse-Compton (EIC) $Y$ parameter.  The IC emission is bolometrically more important than the synchrotron emission if $Y_{\rm IC}>1$.  
We expect that the SSC emission is typically weak.  This is because the SSC $Y$ parameter in the Thomson regime~\citep[e.g.,][]{se01},
\begin{equation}
Y_{\rm SSC}\sim\frac{-1+\sqrt{1+4\eta(\epsilon_l/\epsilon_B)(V_s/c)}}{2}\sim\eta(\epsilon_l/\epsilon_B)(V_s/c),
\end{equation}
is less than unity.  Here, $\eta={\rm min}[1,{(\gamma_c/\gamma_i)}^{2-q}]$ and $\epsilon_l=\epsilon_e$ (or $\epsilon_{\pm}$), where $\gamma_i=\gamma_l$ (or $\gamma_h$) is the injection Lorentz factor and $\gamma_c$ is the cooling Lorentz factor defined below.  
External radiation fields are mainly supplied by SN emission, which is more relevant in our cases.  If the system is optically thin, the energy density of thermal photons is
\begin{equation}
U_{\rm rad}\approx\frac{L_{\rm rad}}{4\pi R^2c}\simeq0.027~{\rm erg}~{\rm cm}^{-3}~L_{\rm rad,42}~R_{16}^{-2},
\end{equation}
whereas the magnetic field energy density is
\begin{eqnarray}
U_B&=&\frac{B^2}{8\pi}\simeq0.064~{\rm erg}~{\rm cm}^{-3}~\varepsilon_{B,-2}\nonumber\\
&\times&D_{*,-1}R_{16}^{-2}{(V_s/5000~{\rm km}~{\rm s}^{-1})}^2.
\end{eqnarray}
Because of $U_B>U_{\rm rad}$, we see that the synchrotron cooling would be typically stronger than the EIC cooling, although the situation can be altered depending on parameters such as $\varepsilon_B$.  As a result, non-thermal X rays can be expected mainly due to EIC emission.  But they will be weaker than thermal X rays (except at hard X rays), and this work focuses on radio signals.

In the dense CSM environment, one also has to care about other losses such as bremsstrahlung and Coulomb losses.  The relativistic bremsstrahlung cooling time scale is~\cite{sch02}
\begin{eqnarray}
t_{e-\rm ff}&\approx&\frac{\pi}{3\alpha\sigma_T c n_{\rm cs}(\ln\gamma_e+\ln 2-1/3)}\nonumber\\
&\simeq&5.2\times{10}^7~{\rm s}~D_{*,-1}^{-1}R_{16}^2{(\ln \gamma_{e,2})}^{-1},
\end{eqnarray}
which is longer than $t_{\rm dyn}$, $t_{e-\rm syn}$ and $t_{e-\rm IC}$ for our typical parameters.  The Coulomb loss time scale of relativistic electrons is~\cite{sch02}
\begin{eqnarray}
t_{e-C}&\approx&\frac{\gamma_e}{0.75 c \sigma_T n_e(60+\ln [\gamma_{e,2}/n_{e,8}])}\nonumber\\
&\simeq&2.2\times{10}^8~{\rm s}~\gamma_{e,2}\mu_eD_{*,-1}^{-1}R_{16}^{2}, 
\end{eqnarray}
which suggests that sufficiently high-energy electrons typically cool via the synchrotron emission.  At lower energies, however, the Coulomb loss can be the shortest time scale, and the resulting synchrotron spectrum becomes complicated.  

The cooling Lorentz factor is given by equating $t_{\rm dyn}^{-1}=t_{e-\rm syn}^{-1}+t_{e-\rm IC}^{-1}$ as
\begin{equation}
\gamma_c\approx25~\varepsilon_{B,-2}^{-1}D_{*,-1}^{-1}R_{16}{(V_s/5000~{\rm km}~{\rm s}^{-1})}^{-1}{(1+Y_{\rm IC})}^{-1}.
\end{equation}
If $\gamma_c<\gamma_i$, the system is in the fast cooling regime, so the energy flux has a peak at $\nu_i\approx\gamma_i^2eB/(m_ec)$. 
If $\gamma_i<\gamma_c$, the system is in the slow cooling regime, so the energy flux has a peak at $\nu_c\approx\gamma_c^2eB/(m_ec)$. 

Primary protons with $\gamma_p>1.37$ lead to inelastic $pp$ reactions, providing secondary electrons.  
The characteristic frequency of electrons with $\gamma_h$ is
\begin{equation}
\nu_{h}\sim1.0\times{10}^{11}~{\rm Hz}~\varepsilon_{B,-2}^{1/2}D_{*,-1}^{1/2}R_{16}^{-1}{(V_s/5000~{\rm km}~{\rm s}^{-1})}.
\end{equation}
For our typical parameters, $D_*\sim0.01-1$ and $R\sim{10}^{15}-{10}^{17}$~cm, we expect $\nu_h\sim3-3000$~GHz.  
We here point that these secondary electrons can be more relevant in the interaction-powered SN scenario.  Very naively, the secondary electronic emission dominates over the primary electronic emission when
\begin{equation}
\epsilon_\pm\approx\frac{1}{6}{\rm min}[1,f_{pp}]\epsilon_p>\epsilon_e,
\end{equation}
which is likely to be realized if $f_{pp}$ is as high as $\sim0.1-1$.  One should keep in mind that this is the crude argument applied to the bolometric emission, and the relative importance changes with frequency, depending on spectral energy distributions of CR protons and electrons.  Assuming the fast cooling regime, the resulting synchrotron flux from hadronically-injected electrons (at $\nu>\nu_h$) is
\begin{eqnarray}
\nu F_\nu^h&\approx&\frac{1}{4\pi d^2}\frac{1}{12}{\rm min}[1,f_{pp}]E_p^2\frac{dN_p}{dE_pdt}\nonumber\\
&\sim&2.7\times{10}^{-14}~{\rm erg}~{\rm cm}^{-2}~{\rm s}^{-1}\nonumber\\
&\times&{\rm min}[10,f_{pp,-1}]{\left(\frac{\nu}{\nu_h}\right)}^{2-q_p}\epsilon_{p,-1}\nonumber\\
&\times&D_{*,-1}{(V_s/5000~{\rm km}~{\rm s}^{-1})}^{3}{\mathcal R}_{p0,1}^{-1}{(d/10~{\rm Mpc})}^{-2},
\end{eqnarray}
which corresponds to $F_\nu^h\sim0.03~{\rm Jy}~{(d/10~{\rm Mpc})}^{-2}~\nu_{11}^{-1}$ for $q_p\sim2$.  Hence, the synchrotron signal is detectable with high-frequency radio telescopes when several absorption processes we discuss below are irrelevant.  In particular, the ALMA sensitivity at 100 GHz is $\sim6\times{10}^{-18}~{\rm erg}~{\rm cm}^{-2}~{\rm s}^{-1}$, allowing detections up to $d\sim0.3-1$~Gpc if followup observations are successful.  

Primary electrons can also emit synchrotron photons, and the corresponding characteristic frequency of electrons with $\gamma_l$ is 
\begin{eqnarray}
\nu_{l}&\approx&5.8\times{10}^{8}~{\rm Hz}~\epsilon_{e,-3}^2f_{e,-5}^{-2}{(g_{q_e}/0.2)}^2\varepsilon_{B,-2}^{1/2}\nonumber\\
&\times&D_{*,-1}^{1/2}R_{16}^{-1}{(V_s/5000~{\rm km}~{\rm s}^{-1})}^5.
\end{eqnarray}
It suggests that studying radio SNe at relatively low frequencies can probe electron acceleration in the shock transition layer, whereas investigations at high frequencies allow us to see the conventional shock acceleration of electrons and/or hadronic injections.  The synchrotron energy flux (at $\nu>\nu_l$) in the fast cooling case is 
\begin{eqnarray}
\nu F_\nu^l&\approx&\frac{1}{4\pi d^2}\frac{1}{2}E_e^2\frac{dN_e}{dE_edt}\nonumber\\
&\sim&1.6\times{10}^{-15}~{\rm erg}~{\rm cm}^{-2}~{\rm s}^{-1}~{\left(\frac{\nu}{\nu_l}\right)}^{2-q_e}\epsilon_{e,-3}\nonumber\\
&\times&D_{*,-1}{(V_s/5000~{\rm km}~{\rm s}^{-1})}^{3}{\mathcal R}_{e0,1}^{-1}{(d/10~{\rm Mpc})}^{-2},
\end{eqnarray}
where $q_e$ is the electron spectral index and we expect $F_\nu^l\sim0.002~{\rm Jy}~{(d/10~{\rm Mpc})}^{-2}~\nu_{11}^{-1}$ for $q_e\sim2$.  Note that $q_e$ may naturally be different from values expected in the conventional diffuse shock acceleration theory at $\lesssim2.1\times{10}^{10}~{\rm Hz}~\varepsilon_{B,-2}^{1/2}D_{*,-1}^{1/2}R_{16}^{-1}{(V_s/5000~{\rm km}~{\rm s}^{-1})}^3$, corresponding to electrons with $\gamma_e\lesssim(m_p/m_e)(V_s/c)$, and steep indices of $q_e\sim3$ are indeed indicated in non-relativistic radio SNe.  The synchrotron signal from primary electrons is also detectable by current radio telescopes and mm/submm facilities including ALMA if secondary electrons are sub-dominant.  
For a given $\nu$, the secondary electronic emission dominates over the primary electronic emission when
\begin{equation}
{\rm min}[10,f_{pp,-1}]\frac{\epsilon_{p,-1}}{\epsilon_{e,-3}}{\left(\frac{\nu_l}{\nu_h}\right)}^{2-q}\gtrsim0.06,
\end{equation}
where the fast cooling regime is assumed for $q_p=q_e=q$.  Hence, we expect that secondaries are typically more important for high-frequency radio emission from interaction-powered SNe. 

However, detecting radio signals may suffer from scattering and various absorption processes.  
First, if $\tau_T\gtrsim{\rm a~few}$, the synchrotron emission can be modified by Comptonization due to thermal electrons.  In particular, thermal electrons in the hot downstream may up-scatter low-energy photons.  The condition that the Comptonization does not change the synchrotron spectrum is roughly given by $y_{\rm NR}\approx (4kT_e/m_ec^2){\rm max}[\tau_T,\tau_T^2]\lesssim1$, so we focus on ejecta-CSM collisions satisfying $D_{*,-1}\lesssim19T_{e,8}^{-1/2}\mu_eR_{16}$.  In addition, there are three important absorption processes that can hinder observations at the radio band, Razin-Tsytovich (RT) suppression, synchrotron self-absorption (SSA) and free-free absorption.  We discuss these suppression and absorption processes below.  

\begin{figure}
\includegraphics[width=\linewidth]{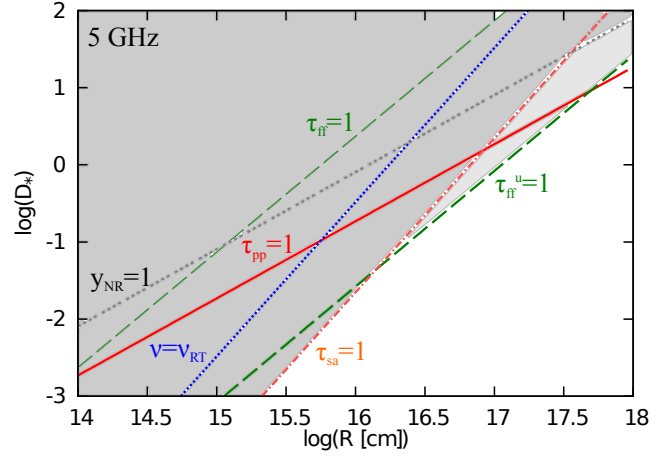}
\caption{The parameter range allowing radio emission in the ($R$, $D$) plane, at $\nu=5$~GHz, where $V_s=5000~{\rm km}~{\rm s}^{-1}$, $\epsilon_B={10}^{-2}$ and $\tilde{f}={10}^{-2.5}$ and $\gamma_i=\gamma_h$ are used.  The suppression or absorption of radio emission is insignificant at $\tau_{\rm sa}<1$, $\nu>\nu_{\rm RT}$, $\tau_{\rm ff}<1$ and $\tau_{\rm ff}^u<1$.  The light shaded region indicates the forbidden region for the radio emission in the pessimistic case, where the upstream is assumed to be ionized with $T_e^u={10}^{5}$~K.  The dark shaded area indicates the forbidden region in the optimistic case, which may be realized for different upstream properties.  The downstream temperature is set to $kT_e=50$~keV.  From this figure, one sees that radio emission from interaction-powered SNe satisfying $\tau_{pp}\sim1$ is suppressed at this band. 
}
\end{figure}

\begin{figure}
\includegraphics[width=\linewidth]{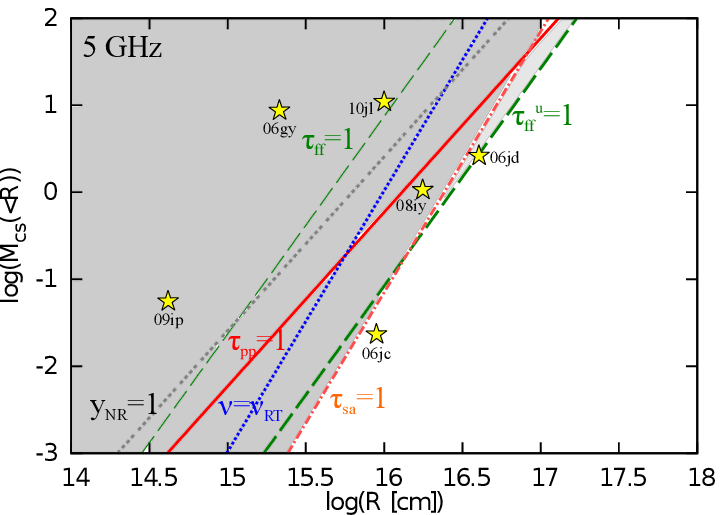}
\caption{The same as Figure~6, but for the ($R$, $M_{\rm cs}$) plane, where $M_{\rm cs}$ has the solar-mass unit.
}
\end{figure}
\begin{figure}
\includegraphics[width=\linewidth]{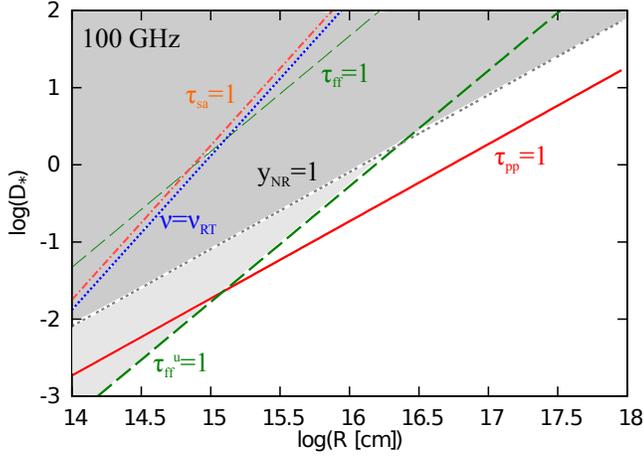}
\caption{The same as Figure~6, but at $\nu=100$~GHz.  From this figure, one sees that observations at 100~GHz are useful for probing interaction-powered SNe satisfying $\tau_{pp}\sim1$.
}
\end{figure}

\begin{figure}
\includegraphics[width=\linewidth]{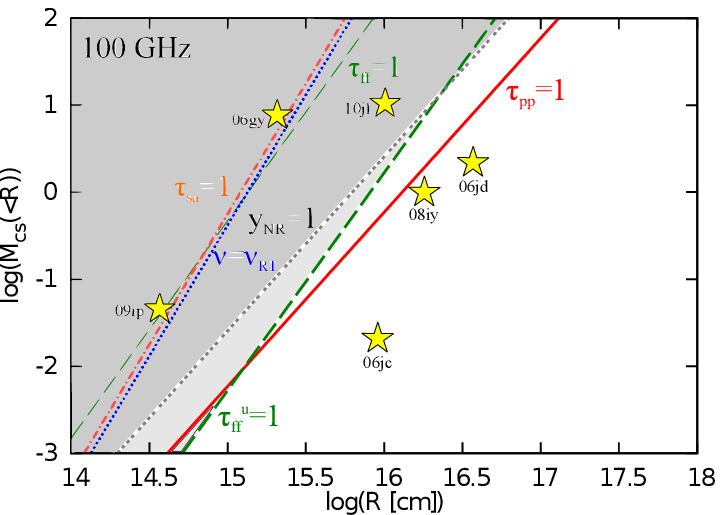}
\caption{The same as Figure~8, but for the ($R$, $M_{\rm cs}$) plane, where $M_{\rm cs}$ has the solar-mass unit.
}
\end{figure}
\begin{figure}
\includegraphics[width=\linewidth]{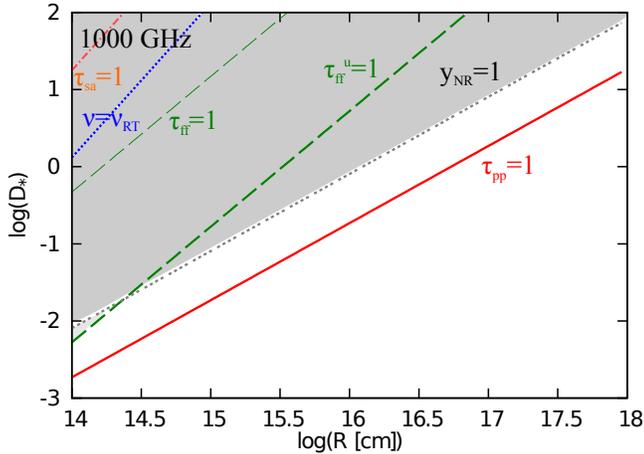}
\caption{The same as Figure~6, but at $\nu=1000$~GHz.
}
\end{figure}

\begin{figure}
\includegraphics[width=\linewidth]{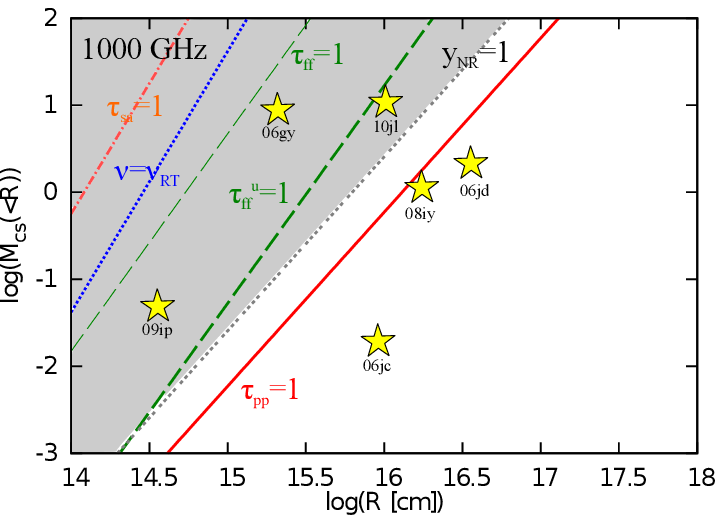}
\caption{The same as Figure~10, but for the ($R$, $M_{\rm cs}$) plane, where $M_{\rm cs}$ has the solar-mass unit.
}
\end{figure}

\subsection{Razin-Tsytovich suppression}
Synchrotron emission in a plasma is different from that in a vacuum.  When a cold plasma plays a role\footnote{This may not be true in a relativistic plasma.}, it is suppressed at low frequencies due to collective effects.  The suppression occurs at the RT frequency~\cite{rl86},
\begin{eqnarray}
\nu_{\rm RT}&\equiv&\frac{2ec n_e}{B}\nonumber\\
&\simeq&8.6\times{10}^9~{\rm Hz}~\varepsilon_{B,-2}^{-1/2}\mu_e^{-1}D_{*,-1}^{1/2}\nonumber\\
&\times&{(V_s/5000~{\rm km}~{\rm s}^{-1})}^{-1}R_{16}^{-1}.
\end{eqnarray}
The line of $\nu=\nu_{\rm RT}$ is shown in Figures~6-11.  Instead, given a frequency, we can obtain the upper limit on $D_*$ as
\begin{equation}
D_{*,-1}\lesssim1.3\times{10}^3~\mu_{e}^{2}\varepsilon_{B,-2}R_{16}^{2}{(V_s/5000~{\rm km}~{\rm s}^{-1})}^{2}\nu_{11}^2.
\end{equation}

\subsection{Synchrotron self-absorption}
The SSA opacity is estimated to be (see Appendix B)
\begin{equation}
\tau_{\rm sa}(\nu)\approx\tau_{\rm sa0} 
\left\{ \begin{array}{ll} 
{(\nu/\nu_n)}^{-\frac{5}{3}},
& \mbox{($\nu<\nu_n$)}\\
{(\nu/\nu_n)}^{-\frac{p+4}{2}}
& \mbox{($\nu_n\leq\nu$)}
\end{array} \right.
\end{equation}
where $p$ is the spectral index of electrons (that is generally different from $q$ for injected particles), $\nu_n\approx\gamma_n^2eB/(m_ec)$, and  
\begin{equation}
\tau_{\rm sa0}=\xi_p\frac{en_{{\rm CR}e}}{B\gamma_n^5}R.
\end{equation}
Here, when the Coulomb cooling is irrelevant, $\gamma_n={\rm min}[\gamma_i,\gamma_c]$ and $\xi_p$ is evaluated as a function of $p$, which is order of $\sim5-10$ (see Appendix B).  When only synchrotron and IC (in the Thomson regime) losses are relevant, we have $p=2$ at $\gamma_c<\gamma_e<\gamma_i$ in the fast cooling case or $p=q$ when $\gamma_i<\gamma_e<\gamma_c$ (i.e., the slow cooling case).  Here, $n_{{\rm CR}e}$ is related to the number of electrons swept by the shock as
\begin{equation}
N_{{\rm CR}e}=4\pi R^3n_{{\rm CR}e}\equiv\tilde{f}\frac{\mathcal E}{m_pc^2},
\end{equation}
where $\tilde{f}$ is the number fraction of electrons defined against $\mathcal E/(m_pc^2)$.  Note that, for primary electron acceleration, the different parameter $f_e$ satisfies
\begin{equation}
N_{{\rm CR}e}=f_e\frac{4\pi DR}{m_H}.
\end{equation}  
Assuming a power-law injection spectrum, we have 
\begin{equation}
\tilde{f}_e=\frac{\epsilon_e}{\gamma_l\ln(\gamma_e^M/\gamma_l)}\frac{m_p}{m_e},
\end{equation}
for $q=2$ and
\begin{equation}
\tilde{f}_e=\frac{\epsilon_e(q-2)}{\gamma_l(q-1)}\frac{m_p}{m_e},
\end{equation}
for $q>2$, and we typically expect $\tilde{f}_e\sim3\times{10}^{-3}\epsilon_{e,-3}$.  For secondary electrons, we have 
\begin{equation}
\tilde{f}_\pm=\frac{{\rm min}[1,f_{pp}]\epsilon_p}{6\gamma_h\ln(\gamma_p^M)}\frac{m_p}{m_e},
\end{equation}
for $q=2$ and
\begin{equation}
\tilde{f}_\pm=\frac{{\rm min}[1,f_{pp}]\epsilon_p(q-2)}{6\gamma_h(q-1)}\frac{m_p}{m_e},
\end{equation}
for $q>2$.  The proton spectrum is assumed to be a power law above $m_p c^2$, where we typically obtain $\tilde{f}_\pm\sim3\times{10}^{-3}\epsilon_{p,-1}{\rm min}[10,f_{pp,-1}]$.  Introducing $\tilde{f}$ allows us to discuss primary and secondary electrons in parallel.  

Setting $\tau_{\rm sa}=1$, the SSA frequency is estimated to be
\begin{eqnarray}
\nu_{\rm sa}&\sim&9.0\times{10}^{9}~{\rm Hz}~\tilde{f}_{-2.5}^{1/3}\varepsilon_{B,-2}^{1/3}D_{*,-1}^{2/3}\nonumber\\
&\times&R_{16}^{-1}{(V_s/5000~{\rm km}~{\rm s}^{-1})}^{4/3}\gamma_{n,1.5}^{1/3},
\end{eqnarray}
for $p=2$ (leading to $\xi_p\simeq8.773$), and
\begin{eqnarray}
\nu_{\rm sa}&\sim&1.1\times{10}^{10}~{\rm Hz}~\tilde{f}_{-2.5}^{2/7}\varepsilon_{B,-2}^{5/14}D_{*,-1}^{9/14}\nonumber\\
&\times&R_{16}^{-1}{(V_s/5000~{\rm km}~{\rm s}^{-1})}^{9/7}\gamma_{n,1.5}^{4/7},
\end{eqnarray}
for $p=3$ (leading to $\xi_p\simeq26.31$).
Instead, given a frequency, we can obtain the upper limit on $D_*$ as
\begin{eqnarray}
D_{*,-1}&\lesssim&37~\tilde{f}_{-2.5}^{-1/2}\varepsilon_{B,-2}^{-1/2}R_{16}^{3/2}\nonumber\\
&\times&{(V_s/5000~{\rm km}~{\rm s}^{-1})}^{-2}\gamma_{n,1.5}^{-1/2}\nu_{11}^{3/2},
\end{eqnarray}
for $p=2$, and
\begin{eqnarray}
D_{*,-1}&\lesssim&32~\tilde{f}_{-2.5}^{-4/9}\varepsilon_{B,-2}^{-5/9}R_{16}^{14/9}\nonumber\\
&\times&{(V_s/5000~{\rm km}~{\rm s}^{-1})}^{-2}\gamma_{n,1.5}^{-8/9}\nu_{11}^{14/9},
\end{eqnarray}
for $p=3$.  Note that we should use $\gamma_n=\gamma_c\propto D^{-1}R$ and $p=2$ for the fast cooling case.  Results for $q=2$ and $q=3$ are shown in Figures~6-11.  In this work, we assume the case of $\nu_{\rm sa}<\nu_b={\rm max}[\nu_c,\nu_i]$, although SSA heating is relevant when $\nu_{\rm sa}>\nu_b={\rm max}[\nu_c,\nu_i]$~\cite{mur+14}.

\subsection{Free-free absorption}
The free-free absorption is important especially when photons propagate in the ionized plasma.  For simplicity, here we assume that ions are protons.  In the hot downstream, the free-free opacity for photons with $h\nu<kT_e$ is~\cite{rl86} 
\begin{eqnarray}
\tau_{\rm ff}(\nu)&\approx&\sigma\alpha_{ff}R\nonumber\\
&\simeq&6.4\times{10}^{-5}~\bar{g}_{\rm ff}T_{e,8}^{-3/2}\mu_e^{-1}D_{*,-1}^2R_{16}^{-3}\nu_{11}^{-2},
\end{eqnarray}
where $\alpha_{\rm ff}$ is the free-free absorption coefficient and $\bar{g}_{\rm ff}$ is the Gaunt factor.  High temperatures of $T_e\sim{10}^{8}$~K are expected in the immediate downstream, while the temperature is lower at the far downstream due to bremsstrahlung cooling especially if the shock is radiative.  The free-free absorption frequency is given by $\tau_{\rm ff}=1$, and we have 
\begin{equation}
\nu_{\rm ff}\simeq8.0\times{10}^{8}~{\rm Hz}~\bar{g}_{\rm ff}^{1/2}T_{e,8}^{-3/4}\mu_e^{-1/2}D_{*,-1}R_{16}^{-3/2}. 
\end{equation}
Instead, given a frequency, we can obtain the upper limit on $D_*$ as
\begin{equation}
D_{*,-1}\lesssim130~\bar{g}_{\rm ff}^{1/2}T_{e,8}^{3/4}\mu_e^{1/2}R_{16}^{3/2}\nu_{11}.
\end{equation}
If the free-free absorption in the emission region is dominant, we expect the suppression factor of $1/\tau_{\rm ff}\propto \nu^2$ at $\nu<\nu_{\rm ff}$.

However, one typically expects that upstream material would be more crucial for absorbing low-frequency emission~\citep[but see][]{cha+12}.  The upstream temperature is lower than the immediate downstream temperature, so the free-free optical depth can be much larger.  Before the shock reaches $R_w$, assuming ionized material, the free-free optical depth in the upstream is 
\begin{eqnarray}
\tau_{\rm ff}^u(\nu)&\approx&\alpha_{ff}R\nonumber\\
&\simeq&5.0\times{10}^{-1}~\bar{g}_{\rm ff}{(T_{e,5}^u)}^{-3/2}\mu_e^{-1}D_{*,-1}^2R_{16}^{-3}\nu_{11}^{-2},
\end{eqnarray}
The free-free absorption frequency is given by $\tau_{\rm ff}^u=1$, and we have 
\begin{equation}
\nu_{\rm ff}\simeq7.1\times{10}^{10}~{\rm Hz}~\bar{g}_{\rm ff}^{1/2}{(T_{e,5}^u)}^{-3/4}\mu_e^{-1/2}D_{*,-1}R_{16}^{-3/2}. 
\end{equation}
Instead, given a frequency, we can obtain the upper limit on $D_*$ as
\begin{equation}
D_{*,-1}\lesssim1.4~\bar{g}_{\rm ff}^{1/2}{(T_{e,5}^u)}^{3/4}\mu_e^{1/2}R_{16}^{3/2}\nu_{11}.
\end{equation}
Note that, if the free-free absorption in the screen region is dominant, we expect the suppression of $\exp(-\tau_{\rm ff}^u)$, which can be in principle distinguished from the other possibilities.  
As shown in Figures~8 and 9, at $\sim100$~GHz, the free-free absorption is typically the most important attenuation process in SNe with dense CSM.  Even at $\sim5$~GHz, it is the dominant attenuation process for ejecta-CSM interactions at $\gtrsim{10}^{16}$~cm.  This process is sensitive to the electron temperature, and $T_e^u={10}^5$~K is used in Figures~6-11.  The temperature may indeed be high enough, as suggested by successful radio detections of SN 2006jd~\cite{cha+12}.  In Figure 7, SN 2006jd~\footnote{While SN 2006jd lies around the light shaded area in Figure 7, the observed radio spectrum may not be consistent with the free-free absorption in the screen zone~\cite{cha+12}.} lies near the lines of $\tau_{\rm ff}^u=1$ and $\tau_{\rm sa}=1$.  On the other hand, non-detections of radio emission from many other Type IIn SNe like SN 2010jl seem consistent with the large absorption that is easily realized with more conservative values of $T_e^u\sim{10}^4$~K~\cite{cha+12b,ofe+13d}.  

At $\sim5$~GHz, as shown in Figures~6 and 7, the free-free absorption and SSA processes suppress radio signals, and the parameter space around $\tau_{pp}\sim1$ is located in the dark shaded area.  So it is difficult to see hadronic signatures with synchrotron emission at radio wavelengths.  But the situation drastically changes at higher frequencies.  At $\sim100$~GHz, the free-free absorption is still an obstacle for ejecta-CSM interactions at $\lesssim{10}^{16}$~cm, but not for $\gtrsim{10}^{16}$~cm.  Importantly, as shown in Figures~8 and 9, a large parameter space of $\tau_{pp}\sim1$ is free of absorption and scattering processes.  So observations at high-frequency radio wavelengths including mm/submm and FIR bands are indeed powerful to test the hadronic model and probe cosmic-ray proton acceleration.  At higher frequencies such as 1000~GHz, all the absorption processes discussed here are negligible compared to the Comptonization due to thermal electrons.  Not all interaction-powered SNe allow us to expect high-frequency radio signals from secondaries.  As pointed out by Murase et al. (2011), except at sufficiently late phases, optically-bright SLSNe are difficult to detect with synchrotron emission at radio bands.  On the other hand, as shown in Figures 9 and 11, some Type IIn SNe such as SN 2006jd and 2008iy seem very promising.        
 
There are some possibilities that we have effectively lower values of $\tau_{\rm ff}^u$.  First, the CSM may be anisotropic or clumpy as suggested in several Type IIn SNe like SN 2005ip~\cite{smi+09} and 2009ip~\citep[e.g.,][]{mar+13}, where the emission more easily escape from partial regions where the CSM density is much lower.  Secondly, $\tau_{\rm ff}^u$ is smaller when the CSM is little ionized, which may be realized especially in the far upstream.  This is different from soft X rays that are more strongly absorbed in the non-ionized CSM.  Such more transparent cases correspond to the light shaded area in Figures~6-11.  

Note that the free-free optical depth declines after the shock reaches the outer edge of CSM, $R_w$.  After $R\gtrsim R_w$, it becomes
\begin{eqnarray}
\tau_{\rm ff}^u(\nu)&=&\int_{R_w}dr\,\,\alpha_{\rm ff}\nonumber\\
&\simeq&5.0\times{10}^{-2}~\left(\frac{10}{2s'-1}\right)\bar{g}_{\rm ff}{(T_{e,5}^u)}^{-3/2}\mu_e^{-1}D_{*,-1}^2\nonumber\\
&\times&R_{w,16}^{-3}{(R_w/R)}^{2s'-1}\nu_{11}^{-2},
\end{eqnarray}
where $\varrho_{\rm cs}\propto R^{-s'}$ (at $R>R_w$) is assumed.  For example, a possible value of $s'\sim5-6$ is suggested in the late phase of the re-brightening of SN 2009ip~\cite{mar+13}.

\section[]{Summary}
In this work, we provide a broad discussion of multi-messenger diagnoses of interaction-powered SNe including Type IIn SNe and some SLSNe, focusing on non-thermal signals.  The shock would be radiation-mediated at very early times, and thus CR acceleration is inefficient.  However, as photons escape the system, a collisionless shock can form and CR acceleration becomes possible.  While shock heating leads to X rays, CRs are expected to produce broadband non-thermal emission, including gamma rays, X rays, radio waves and neutrinos.  

Photon emission in general may be largely thermalized depending the optical depth, which in turn depends on details of the CSM.  Neutrinos are the most direct probe in the sense that they do not suffer from attenuation in the source.  In addition, by advancing the idea proposed by Murase et al. (2011), we have shown that GeV gamma rays can typically escape after the shock breakout, although TeV gamma rays are attenuated due to the two-photon annihilation process~\footnote{Note that Model A in Murase et al. (2011) considered a different situation before breakout of the forward shock emission.}.  Along with neutrinos, GeV gamma rays can provide unique opportunities to probe the formation of collisionless shocks and the onset of CR acceleration.  Interestingly, the physical parameters suggested by observed interaction-powered SNe imply densities similar to those inferred from gamma-ray novae~\cite{abd+10}, allowing us to expect analogous high-energy emission and to probe the physics of CR acceleration in the dense environment.  Detecting signals from one SN requires a nearby event, but stacking analyses are still useful.  Gamma rays and neutrinos are especially powerful for optically-bright SLSNe, for which the recipes provided in Appendix A can be used.     

For normal luminosity interaction-powered SNe, broad-band non-thermal emissions from radio to TeV gamma-ray bands are possible.  In particular, high-frequency radio observations in the mm/submm and FIR bands can probe CR proton acceleration and test the hadronic model.  We pointed out that secondary electrons produced via $pp$ reactions play an important pole role in the synchrotron emission from some interaction-powered SNe such as SN 2006jd.  Comprehensive observations from GHz to 1000~GHz may also be relevant to study acceleration of primary electrons that may not be accelerated by the conventional shock acceleration.  Our work demonstrates the importance of multi-messenger approaches in revealing the mechanism of Type IIn SNe and CR acceleration in real time. 

The interaction-powered SN scenario has been commonly used to interpret SLSNe, but SLSNe are diverse and other scenarios also possible.  For example, some SLSNe such as SN 2007bi may be rather pair-instability SNe originating from progenitors with $M_*\gtrsim130 M_\odot$, where the stellar collapse is caused by the pressure decrease due to electron-positron pair-production~\cite{gl09,gal+09}.  Or luminous SNe including SLSNe Ic~\citep[e.g.,][]{cho+13} may be driven by newborn pulsars~\citep[e.g.,][]{kb10,met+11,woo10}.  Some SLSNe seem difficult to explain using these scenarios~\cite{gal12,ins+13}.  High-energy emissions including gamma rays and neutrinos have been predicted in both the interaction-powered SN scenario~\cite{mur+11,kat+11} and the pulsar-driven SN scenario~\cite{mur+09,kot+13}.  Detecting thermal and non-thermal signals from shocks in the dense CSM and studying time-dependent spectra are crucial in order to discriminate among the scenarios. 

The most important point of this work is that secondary electrons and positrons from inelastic $pp$ collisions will radiate detectable synchrotron photons efficiently at high-frequency radio wavelengths including the mm/submm and FIR bands.  Although details depend on the mass of CSM, its physical location relative to the progenitor star at the time of explosion, and the velocity of the ejecta, for typical parameters, we expect the synchrotron spectrum to peak at $\sim3-3000$~GHz and with flux of $\sim0.01-0.1$~mJy for 2006jd-like interaction-powered SNe at distances of hundreds of Mpc.  In particular, high-frequency radio signals using instruments like the high-frequency channels of the VLA and ALMA can be very powerful to probe physics of collisionless shocks.  For this reason, we encourage followup observations especially at the mm/submm band within months-to-years and at the GeV-TeV gamma-ray band within days-to-years.

\section*{Acknowledgments}
K. M. thanks John Beacom, Roger Chevalier, Boaz Katz, Brian Lacki, Keiichi Maeda, Jose Prieto, Masaomi Tanaka and Ryo Yamazaki for useful discussions.
K. M. is supported by NASA through Hubble Fellowship, Grant No. 51310.01 awarded by the Space Telescope Science Institute, which is operated by the Association of Universities for Research in Astronomy, Inc., for NASA, under Contract No. NAS 5-26555.  K. M. also thanks supports by CCAPP since this project was started in the beginning of 2011 when K. M. was a member of CCAPP.  The preliminary results of this work was presented in March 2013, Hubble Fellows Symposium.   

\appendix
\section[]{Recipes for testing breakout high-energy emission in the interaction-powered SN scenario}
Here, we provide recipes to test the hadronic model for gamma-ray and neutrino emissions from interaction-powered SNe.  Basically, we need to know two quantities, the CR energy ${\mathcal E}_{{\rm CR}p}$ and the $pp$ efficiency $f_{pp}$.  The latter depends on $n_{\rm cs}$, $R$ and $V_s$, which can be determined from the breakout emission. 

The more sophisticated approach is possible based on self-similar solutions in the engine-driven case~\cite{che82}.  We here overview the prescription by Chevalier \& Irwin (2011).  Let us consider the ejecta whose outer density profile is $\varrho_{\rm ej}=Ct^{-3}{(R/t)}^{-7}$ at $V_r>V_t={{(2\mathcal E}_{\rm ej}/M_{\rm ej})}^{1/2}$.  The shock is initially radiation-mediated and the flows are radiation-dominated since $\tau_T\gg(c/V_s)$. The contact discontinuity is located at $R_{\rm cd}={(0.227C/D)}^{1/5}t^{4/5}$, where $C=4{\mathcal E}_{\rm ej}^2/(3\pi M_{\rm ej})$.  The forward shock radius is estimated to be
\begin{eqnarray}
R_{f}=1.208R_{\rm cd}&\simeq&0.85\times{10}^{15}~{\rm cm}~{\mathcal E}_{\rm ej,51}^{2/5}{(M_{\rm ej}/{10}^{0.5}~M_\odot)}^{-1/5}\nonumber\\
&\times&D_*^{-1/5}{(t/10~{\rm d})}^{4/5}
\end{eqnarray}
and the forward shock velocity is $V_{f}=(4/5)R_f/t\simeq7800~{\rm km}~{\rm s}^{-1}~{\mathcal E}_{\rm ej,51}^{2/5}{(M_{\rm ej}/{10}^{0.5}~M_\odot)}^{-1/5}D_*^{-1/5}{(t/10~{\rm d})}^{-1/5}$.  The energy carried by the interacting shell is estimated to be ${\mathcal E}\approx4{\mathcal E}_{\rm ej}^2/(3 M_{\rm ej}V_0^2)$, where $V_0\approx R_r/t\simeq7900~{\rm km}~{\rm s}^{-1}~{\mathcal E}_{\rm ej,51}^{2/5}{(M_{\rm ej}/{10}^{0.5}~M_\odot)}^{-1/5}D_*^{-1/5}{(t/10~{\rm d})}^{-1/5}$.  For $\hat{\gamma}=4/3$, the radiation energy is ${\mathcal E}_{\rm rad}=0.32\mathcal E$, so we have
\begin{equation}
{\mathcal E}_{\rm rad}\simeq1.1\times{10}^{50}~{\rm erg}~{\mathcal E}_{\rm ej,51}^{6/5}{(M_{\rm ej}/{10}^{0.5}~M_\odot)}^{-3/5}D_*^{2/5}{(t/10~{\rm d})}^{2/5}.
\end{equation}
In the wind case, the photon diffusion timescale is comparable to the breakout time, and we have
\begin{equation}
t_{\rm rise}\approx\frac{R_{\rm bo}}{V_f}\approx\frac{\sigma_T}{\mu_em_H}\frac{D}{c}\simeq7.7~{\rm d}~\mu_e^{-1}D_*,
\end{equation}
as long as the breakout happens at $R_{\rm bo}\ll R_w$.  The breakout radius can also be estimated from the evolution of the radiation luminosity and temperature if the black-body approximation is valid, or if $V_f$ is known. 

As a result, if we can observe ${\mathcal E}_{\rm rad}$, $t_{\rm rise}$ and $V_f$ (or $R_{\rm bo}$), we can evaluate $D$, ${\mathcal E}_{\rm ej}^2/M_{\rm ej}$ and $\mu_e$ as~\cite{mar+13} 
\begin{equation}
D_*\simeq2.9~{\mathcal E}_{\rm rad,51}{(t_{\rm rise}/10~{\rm d})}^2R_{\rm bo,15}^{-3},
\end{equation}
\begin{equation}
{\mathcal E}_{\rm ej,51}^2/(M_{\rm ej}/{10}^{0.5}~M_\odot)\simeq17~{\mathcal E}_{\rm rad,51}{(t_{\rm rise}/10~{\rm d})}^{-2}R_{\rm bo,15}^{2},
\end{equation}
\begin{equation}
\mu_e^{-1}\simeq0.45~{\mathcal E}_{\rm rad,51}^{-1}{(t_{\rm rise}/10~{\rm d})}^{-1}R_{\rm bo,15}^{3}.
\end{equation}
One should keep in mind that the engine-driven self-similar solution is valid as long as $V_{\rm r}>V_t$.

Then, we can estimate $n_{\rm cs}$, allowing us to calculate $f_{pp}$ and resulting neutrino and gamma-ray spectra.  
There are two important free parameters, the CR spectral index ($q_p$) and CR energy (${\mathcal E}_{{\rm CR}p}$).  For $q_p=2$, the CR proton spectrum is given by  
\begin{equation}
E_p^2 \frac{dN_p}{dE_p}\equiv\frac{{\mathcal E}_{{\rm CR}p}}{{\mathcal R}_p}=\frac{{\mathcal E}_{{\rm CR}p}}{\ln(E_p^M/E_p^m)}
\end{equation}
and ${\mathcal R}_p\sim15$ for our typical parameters.  For $q_p>2$, we have
\begin{equation}
E_p^2 \frac{dN_p}{dE_p}\equiv\frac{{\mathcal E}_{{\rm CR}p}}{{\mathcal R}_p(E_p)}=(q_p-2){\left(\frac{E_p}{E_p^m}\right)}^{2-q_p}{\mathcal E}_{{\rm CR}p}.
\end{equation}
Here, ${\mathcal R}_p$ is the conversion factor from the total energy to the differential energy spectrum and $E_p^m\sim m_pc^2$ is the minimum proton energy.  These equations can be rewritten as
\begin{equation}
E_p^2 \frac{dN_p}{dE_p}\equiv{\mathcal R}_{p0}^{-1}{\left(\frac{E_p}{E_p^m}\right)}^{2-q_p}{\mathcal E}_{{\rm CR}p},
\end{equation}
where ${\mathcal R}_{p0}\equiv{\mathcal R}_{p}(E_p^m)$.  The CR energy is parametrized as
\begin{equation}
{\mathcal E}_{{\rm CR}p}=\epsilon_p{\mathcal E}=(\epsilon_p/\epsilon_\gamma){\mathcal E}_{\rm rad},
\end{equation}
where $\epsilon_p/\epsilon_\gamma$ is the CR loading parameter that is commonly introduced in the literature of hadronic emissions from gamma-ray bursts~\cite{mn06}.  Since both $\epsilon_p$ and $\epsilon_\gamma$ are order of 0.1, we expect $\epsilon_p/\epsilon_\gamma\sim1$ and we can make predictions for breakout high-energy emissions, based on observational quantities.  Such an application was done in Margutti et al. (2013).  Note that high-energy emissions continue after the breakout.  Given sufficient time-dependent data, later contributions can easily be taken into account by more detailed modeling.  For example, one can directly use self-similar solutions for $\hat{\gamma}=4/3$ or $\hat{\gamma}=5/3$ at $\tau_T\lesssim c/V_s$~\cite{ci11,ofe+13d}.  Most naively, instead, the overall contribution can be incorporated in the CR loading parameter.

\section[]{Formulas of synchrotron self-absorption}
Here we provide formulas to calculate SSA.  For a power-law electron distribution, the SSA coefficient is~\cite{rl86}
\begin{eqnarray}
\alpha_{\rm sa}(\nu)&=&{\mathcal N}_0\frac{\sqrt{3}e^3}{8\pi m_e}{\left(\frac{3e}{2\pi m_e^3c^5}\right)}^{p/2}{(B\sin\theta_B)}^{(p+2)/2}\nonumber\\
&\times&\Gamma(p/4+11/6)\Gamma(p/4+1/6)\nu^{-(p+4)/2},
\end{eqnarray}
where $\theta_B$ is the angle between the electron velocity and magnetic field.  The normalization is determined by
\begin{equation}
\int_{\gamma_n}d\gamma_e\,\,\,{\mathcal N}_0\gamma_e^{-p}=n_{{\rm CR}e}.
\end{equation}
Assuming $\gamma_b={\rm max}[\gamma_i,\gamma_c]\gg\gamma_n={\rm min}[\gamma_i,\gamma_c]$, we have ${\mathcal N}_0\approx\tilde{\xi}_pn_{{\rm CR}e}\gamma_n^{p-1}$, where $\tilde{\xi}_p=p-1$ for $p>1$ and $\tilde{\xi}_p=\ln(\gamma_b/\gamma_n)$ for $p=1$.  
When only the synchrotron emission is relevant, we have $p=2$ at $\gamma_c<\gamma_e<\gamma_i$ in the fast cooling case or $p=q$ at $\gamma_i<\gamma_e<\gamma_c$ in the slow cooling case.  
Averaged over $\theta_B$, the SSA coefficient is written as
\begin{equation}
\alpha_{\rm sa}(\nu)=\xi_p\frac{en_{{\rm CR}e}}{B\gamma_n^5}{(\nu/\nu_n)}^{-(p+4)/2},
\end{equation}
where
\begin{eqnarray}
\xi_p&=&\tilde{\xi}_p \frac{\pi^{3/2} 3^{(p+1)/2}}{4}\Gamma(p/4+11/6)\Gamma(p/4+1/6)\nonumber\\
&\times&\frac{\Gamma (p/4+3/2)}{\Gamma (p/4+2)}.
\end{eqnarray}
For example, we obtain $\xi_2\simeq8.773$ for $p=2$ and $\xi_3\simeq26.31$ for $p=3$, respectively. 

Note that the spectrum in the optically-thick limit is obtained from $F_\nu=\pi(j_{\nu}^{\rm syn}/\alpha_{\rm sa})(R^2/d^2)\propto\nu^{5/2}$, where $j_\nu^{\rm syn}$ is the synchrotron emissivity~\cite{rl86}.  The coefficient agrees with Katz (2012).  One should keep in mind that the $F_{\nu}\propto\nu^{5/2}$ is obtained only if $\nu_{\rm sa}>\nu_n$, whereas we expect $F_{\nu}\propto\nu^{2}$ if $\nu_{\rm sa}<\nu_n$.

\bsp

\label{lastpage}

\end{document}